\pdfoutput=1 
\documentclass[12pt]{article}

\usepackage{inputenc}

\usepackage{amsmath,amssymb,bbm,xcolor}
\usepackage{amsbsy}
\usepackage{fixmath}
\usepackage{euscript}
\usepackage{graphicx}
\usepackage{hyperref}
\usepackage{cite}
\usepackage{xspace}
\usepackage{soul}
\numberwithin{equation}{section}

\newcommand{\msbar}{\overline{\text{MS}}}

\newcommand{\order}[1]{\mathcal{O}\!\left(#1\right)}

\newcommand{\qbar}{{\bar q}}

\newcommand{\as}{{\alpha_s}}

\newcommand{\ncubed}{\text{N$^3$LO}\xspace}

\newcommand{\ie}{{\it i.e. }}

\newcommand{\krknlo}{{\textsf{KrkNLO}}}

\newcommand{\herwig}[1]{\textsf{Herwig~7#1}}
\newcommand{\sherpa}[1]{\textsf{Sherpa#1}}
\newcommand{\mcatnlo}[1]{\textsf{MC@NLO#1}}
\newcommand{\mcfm}[1]{\textsf{MCFM#1}}
\newcommand{\powheg}[1]{\textsf{POWHEG#1}}

\newcommand{\dynnlo}[1]{\textsf{DYNNLO#1}}
\newcommand{\hnnlo}[1]{\textsf{HNNLO#1}}
\newcommand{\LHAPDF}[1]{\textsf{LHAPDF#1}}
\newcommand{\hqt}[1]{\textsf{HqT#1}}

\begin{document}

\begin{titlepage}


\begin{flushright}
\bf IFJPAN-IV-2016-10\\
    CERN-PH-TH-2016-166 \\
    MAN/HEP/2016/13\\
    KA-TP-23-2016\\
    MCnet-16-27\\

\end{flushright}

\vspace{3mm}
\begin{center}
    {\Large\bf Monte Carlo simulations of Higgs-boson production 
    \vspace{2mm} \\
               at the LHC  with the KrkNLO method$^{\star}$ }

\end{center}

\vskip 5mm
\begin{center}
{\large 
           S.\ Jadach$^a$, G.\ Nail$^{b,c}$, W.\ P\l{}aczek$^d$, 
	   S.\ Sapeta$^{e,a}$,  \vspace{2mm} \\
           A.\ Si\'odmok$^{e,a}$ and M.\ Skrzypek$^a$
}

\vskip 6mm
{\em $^a$Institute of Nuclear Physics, Polish Academy of Sciences,\\
  ul.\ Radzikowskiego 152, 31-342 Krak\'ow, Poland} 
\vspace{2mm}\\
{\em $^b$Particle Physics Group, School of Physics and Astronomy, 
           University of Manchester, \\
Oxford Road, Manchester, M13 9PL, United Kingdom}
\vspace{2mm}\\
{\em $^c$Institute for Theoretical Physics, Karlsruhe Institute of Technology, \\
Wolfgang-Gaede-Strasse 1, 76131 Karlsruhe, Germany}
\vspace{2mm}\\
{\em $^d$Marian Smoluchowski Institute of Physics, Jagiellonian University,\\
ul.\ {\L}ojasiewicza 11, 30-348 Krak\'ow, Poland}
\vspace{2mm}\\
{\em $^e$Theoretical Physics Department, CERN, CH-1211 Geneva 23, Switzerland }
\end{center}

\begin{abstract}
\noindent
We present numerical tests and predictions of the \krknlo\ method for matching
of NLO QCD corrections to hard processes with LO parton shower Monte Carlo generators (NLO+PS). 
This method was described in detail in our previous publications, where
it was also compared with other NLO+PS matching approaches
(\mcatnlo{} and \powheg{}) as well as fixed-order NLO and NNLO calculations.
Here we concentrate on presenting some numerical results
(cross sections and distributions) for $Z/\gamma^*$ (Drell--Yan) 
and Higgs-boson production processes at the LHC. 
The Drell--Yan process is used mainly to validate the \krknlo\ 
implementation in the \herwig{} program 
with respect to the previous implementation in \sherpa.
We also show predictions for this process with the new, 
complete, MC-scheme parton distribution functions and compare 
them with our previously published results. 
Then, we present the first results of the \krknlo\ method
for Higgs production in gluon--gluon fusion at the LHC 
and compare them with 
\mcatnlo{} and \powheg{} predictions from \herwig{}, fixed-order results from
\hnnlo{} and a resummed calculation from \hqt{},
as well as with experimental data from the ATLAS collaboration.
\end{abstract}

\vspace{5mm}
\footnoterule
\noindent
{\footnotesize
$^{\star}$This work is partly supported by 
 the Polish National Science Centre grant UMO-2012/04/M/ST2/00240.
}

\end{titlepage}
\section{Introduction}
\label{sec:intro}

The discovery of the Higgs boson, at the Large Hadron
Collider~(LHC)~\cite{Aad:2012tfa,Chatrchyan:2012xdj}, opened a new era in the
exploration of the electroweak sector of the Standard Model~(SM).
The measured value of the Higgs mass uniquely specifies all 
of the couplings and turns the SM into a fully predictive theory.
Hence, we are at a position to perform stringent tests of our current
modelling of these fundamental interactions.
This is only possible if we are in possession of precise theoretical
predictions for the Higgs production cross sections.

Most of the Higgs boson particles observed at hadron colliders are produced
through the process of gluon fusion, a channel that is known to exhibit very
slow convergence in perturbative Quantum Chromodynamics~(QCD).
At LHC energies, the next-to-leading order~(NLO) corrections to the total
cross section for the inclusive production of the Higgs boson through gluon fusion
turn out to be as large as 70\%, and the next-to-next-to-leading order~(NNLO)
corrections increase the cross section by another~30\%~\cite{Harlander:2002wh,
Anastasiou:2002yz, Catani:2007vq}.
The theoretical uncertainty of the NNLO result, arising from the missing higher
orders and obtained by the standard renormalization and factorization scale
variations, is estimated at around 10\%, and is hence at the level of the experimental
accuracy of the Run 1 LHC measurements.
This large uncertainty at NNLO has motivated the efforts to further improve the
precision by calculating the full
next-to-next-to-next-to-leading order~(\ncubed) result for inclusive Higgs
boson production in gluon fusion~\cite{Anastasiou:2015ema}.
Adding these contributions to the predictions for the cross section reduces
their scale uncertainties down to the level of~3\%.
 
Apart from the inclusive Higgs cross section, which is the most fundamental
quantity, as it enables one to predict the total number of Higgs particles produced
at a given energy and luminosity, one is also equally interested in more differential
observables.
Therefore,
a significant amount of work has also gone into obtaining predictions for
differential cross sections for Higgs production in gluon fusion beyond NLO.

In particular,
differential observables have been predicted within frameworks of
analytic resummation, like for example small-$q_T$ resummation performed 
in QCD in coordinate space up to the NNLL+NLO accuracy~\cite{deFlorian:2011xf}
and directly in momentum space up to NNLL+NNLO~\cite{Monni:2016ktx} as well as
within SCET~\cite{Becher:2012yn} up to NNLL+NLO.

Differential cross sections for Higgs production in gluon fusion have been also
widely studied with approaches in which fixed-order NLO results are matched to parton
shower (NLO+PS) such as the MiNLO
method~\cite{Hamilton:2012np,Hamilton:2012rf}.  Recently, NNLO+PS matched
results were computed with the UN$^2$LOPS technique~\cite{Hoche:2014dla} as well
as with an extended version of
MiNLO~\cite{Hamilton:2013fea,Frederix:2015fyz,Hamilton:2016bfu} combined with
the HNNLO program~\cite{Catani:2007vq, Grazzini:2008tf}.
The current methods of performing
NNLO+PS~\cite{Hamilton:2012rf,Hoeche:2014aia,Alioli:2013hqa,Hoche:2015sya,Alioli:2015toa,Hamilton:2015nsa,Frederix:2015fyz,Hamilton:2016bfu}
represent
clear progress in the matching of fixed-order NNLO QCD calculations with
parton shower Monte Carlos (PSMCs).
The next challenge towards even higher-precision  perturbative QCD
calculations, needed until the end of the LHC era two decades from now, is the
combination of the fully exclusive NNLO corrections for the hard process with
NLO parton shower (NNLO+NLOPS).

In this article, we present NLO+PS predictions for various differential
distributions computed with the \krknlo{} approach~\cite{Jadach:2011cr, Jadach:2015zsq}.
The main advantage of the \krknlo{} method with respect to other methods of
matching the fixed-order NLO calculations with PSMCs (\mcatnlo{} and \powheg{}) 
is its simplicity, which stems from the fact that the entire NLO corrections are
implemented using a simple, positive, multiplicative MC weight in combination with
pre-calculated MC-scheme PDFs. The present work is relevant for the above future
developments in the sense that it presents a simplified method of correcting the
hard process to NLO level in combination with a leading order (LO) parton shower (PS)
that will hopefully pave the way to NNLO hard processes combined with a NLO PSMC;
NLOPS is a parton shower MC implementing the NLO evolution kernels in the fully
exclusive form, thus providing the full set of the soft-collinear counter-terms
for the hard process. Ref.~\cite{Jadach:2013dfd} reviews several feasibility
studies which show that constructing such a NLOPS is, in principle, plausible.
In our opinion, any simplifications of the NLO+PS matching, as in the \krknlo{}
method, will be instrumental in the progression towards more ambitious fully
exclusive NNLO+NLOPS projects.

The \krknlo\ method was first introduced Ref.~\cite{Jadach:2015mza}
for $Z/\gamma^*$ production in hadron collisions (the Drell--Yan process, DY) 
and was also presented in Ref.~\cite{Jadach:2015zsq}. 
These developments followed the initial study in Refs.~\cite{Jadach:2011cr,Jadach:2012vs} 
on the inclusion of fixed-order NLO QCD corrections to the hard process 
in LO PSMC through an appropriate Monte Carlo weight. 
This study was performed
with the use of some dedicated toy-model parton-shower generator 
and for gluonstrahlung from quarks only, albeit 
for two processes: DY production and deep-inelastic electron--proton scattering (DIS).

The first realistic numerical results (total cross sections 
and distributions of $Z/\gamma^*$ transverse momentum 
and rapidity) of the \krknlo{} method, 
based on its implementation in the \sherpa{}~\cite{Gleisberg:2008ta} PSMC, 
were presented in Ref.~\cite{Jadach:2015mza} for the DY process.
The \krknlo{} results were compared with the fixed-order 
NLO predictions of \mcfm{}~\cite{MCFM}
and those of other NLO-PSMC matching methods, 
namely \mcatnlo{}~\cite{Frixione:2002ik,Frixione:2006he} and
\powheg{}~\cite{Nason:2004rx,Frixione:2007vw}, 
as well as with the fixed-order NNLO calculations of \dynnlo{}~\cite{Catani:2009sm}.
A satisfactory agreement with other NLO calculations was found. Moreover, 
for the boson transverse momentum
the agreement with the NNLO predictions was better than for \mcatnlo{} and \powheg{},
which may be explained by effects beyond NLO accounted for in \krknlo{} as a result of 
using the MC factorization scheme and multiplicative virtual+soft-real corrections,
see Ref.~\cite{Jadach:2015mza} for more details.
In that paper the advantages of the \krknlo{} method 
over the \mcatnlo{} and \powheg{} techniques were also discussed.

In Ref.~\cite{Jadach:2015mza} the concept of the Monte Carlo (MC) 
factorization scheme was introduced as a necessary
ingredient of the \krknlo{} method; this was further developed in
Ref.~\cite{Jadach:2016acv}, see also Ref.~\cite{Jadach:2016viv}. 
Appropriate MC-scheme parton distribution functions (PDFs) were defined and
constructed from PDFs in the standard $\msbar$ scheme. 
However, those MC-scheme PDFs included only contributions from
parton--parton transitions that were sufficient 
for the DY process in the NLO approximation -- in the following we denote them
with $\rm MC_{DY}$. 
The complete PDFs in the MC factorization scheme, 
which include all of the LO parton--parton transitions, 
were defined, computed and compared with the $\msbar$ PDFs in Ref.~\cite{Jadach:2016acv}, 
where Higgs-boson production from initial-state gluons was considered. 
In that paper only the values of the total cross section
for the Higgs production at the LHC were shown; 
further results for this process will be presented in this work.
However, before presenting the Higgs results, 
we first validate the implementation of the \krknlo{} method in 
the \herwig{} program with respect to its previous implementation 
in \sherpa{} using the DY process. 
The \herwig{} implementation of the \krknlo{} method will be our basic platform 
for its future developments and applications to other processes. 
Then, we will compare the \krknlo{} results for the DY process 
based on the complete MC-scheme PDFs with
those where the  $\rm MC_{DY}$ PDFs, 
as defined in Ref.~\cite{Jadach:2015mza}, are used.

For the process of Higgs-production in gluon--gluon fusion, 
we present the results for the total cross section 
and the distributions of the Higgs transverse momentum and rapidity at the LHC. 
The predictions of the \krknlo{} method are compared 
with those of the fixed-order NLO and NNLO calculations 
of \hnnlo{}~\cite{Catani:2007vq}, the results 
of the NLO-PSMC matching approaches of
\mcatnlo{}~\cite{Frixione:2006he} and \powheg{}~\cite{Nason:2004rx, Frixione:2007vw},
as well as resummed calculations from \hqt{}~\cite{Bozzi:2005wk,deFlorian:2011xf}. 
We also confront the predictions of all the above matching methods with the LHC data
of the ATLAS experiment~\cite{Aad:2015lha}.

The outline of this paper is the following: In Section~2, 
after describing the set-up for our numerical computations, 
we present and discuss the results of the \krknlo{} method; 
we do this first for the DY process and then for Higgs production 
in gluon--gluon fusion at the LHC. 
Section~3 concludes our work and provides some outlook for future developments.
In Appendix~A we present comparisons of various PDF parametrizations
in the $\msbar$ and MC factorization schemes.

\section{Numerical results}
\label{sec:numres}
\subsection{Set-up}
\label{ssec:setup}

For the numerical evaluation of the cross sections%
\footnote{Unless stated otherwise in the text.} 
at the LHC for the proton--proton collision energy of $\sqrt{s}=8\,$TeV
we have chosen the following set of the Standard Model input parameters:
\begin{eqnarray}\label{eq:pars}
M_Z = 91.1876 \; {\rm GeV}, & \quad & \Gamma_Z =  2.4952  \; {\rm GeV},
\nonumber  \\
M_W = 80.4030 \; {\rm GeV}, & \quad & \Gamma_W = 2.1240 \; {\rm GeV},
\\
M_H = 126 \; {\rm GeV},
& \quad & m_t = 173.2  \; {\rm GeV},
\nonumber  \\
      & \quad & G_{\mu} = 1.16637\times 10^{-5} \; {\rm GeV}^{-2}, 
\nonumber  
\end{eqnarray}
and use the $G_{\mu}$-scheme~\cite{LHC-YR} for the electroweak sector of the Standard Model.
We take the renormalization and
factorization scales to be $\mu_R^2=\mu_{F}^2=M_B^2$, where $M_B$ is the mass
of the $Z$ or Higgs boson for the respective processes. 
In the case of the Drell--Yan process, detector acceptance cuts are imposed 
only on the invariant mass of the final-state lepton pair ($Z/\gamma^*$-boson):
\begin{equation}
50~{\rm GeV}<M_{l\bar{l}}<150~{\rm GeV}.
\label{eq:Mllcut}
\end{equation}
For the Higgs-production process we do not apply any cuts, and for simplicity we
set the Higgs boson to be stable. 
The LO, NLO and NNLO Higgs-production matrix elements
are calculated in the $m_t\rightarrow \infty$ and $m_{q\neq t} = 0$ approximation.
To compute the hadronic cross-section, we
employ $\msbar$ PDFs from the \LHAPDF{} library \cite{Whalley:2005nh} and their MC-scheme
counterparts when using \krknlo{}. The value of $\as$ is chosen to match the
value used in the PDFs. The PDF set used is detailed in the relevant subsection.

\subsection{Drell--Yan process}
\label{ssec:DY}

The results of the \krknlo{} method implemented on top of the \sherpa{} 
PSMC~\cite{Gleisberg:2008ta} 
for the Drell--Yan process were already presented in Ref.~\cite{Jadach:2015mza}.
As with these results, for this process we use the {\tt MSTW2008} LO set of parton
distribution functions~\cite{Martin:2009iq}, which has $\as(M_Z^2)=0.13938690$.
Here we use these \sherpa{} results to validate a new implementation of \krknlo{} 
in the \herwig{} program \cite{Bahr:2008pv,Bellm:2015jjp}. 
This version of the \herwig{} PSMC features a new parton-shower 
algorithm based on the Catani--Seymour (CS) dipole 
\cite{Catani:1996vz} formalism, and is therefore well suited 
to the implementation of the \krknlo{} method.
Details on how to implement the \krknlo{} method in the CS-dipole PSMC 
are given in Ref.~\cite{Jadach:2015mza} for
the DY process and in Ref.~\cite{Jadach:2016acv} for the Higgs-boson production. 
In short, it amounts to replacing the $\msbar$ PDFs with 
the MC-scheme PDFs and applying to each event generated 
by the PSMC an appropriate Monte Carlo weight that introduces 
the NLO QCD correction to a given hard process. 
This weight is positive-definite and can be computed simply 
from information provided in an event record. 

\subsubsection{LO results}
\label{sssec:dylo}

As a basis for this validation, we first compare the LO-level predictions 
from \mcfm{} and \sherpa{} with those of \herwig{}
using an identical choice of parameters (see Section. \ref{ssec:setup}). 
The results for the total cross section presented in Table~\ref{tab:Z-lo-settings-val} show
a very good agreement (within statistical errors) between different programs.

\begin{table}[t]
\centering
\begin{tabular}{|c||c|c|c|}
  \hline
                           & \mcfm{}  & \sherpa{} & \herwig{}  \\
  \hline\hline
  $\sigma_\text{tot}$ [pb] &  $936.9\,(1)$     
                    &  $937.2\,(2)$      &  $ 936.6\,(2)$ \\ 
  \hline
\end{tabular}
\caption{\sf
Values of the total cross section with statistical errors at the Born level for 
the Drell--Yan process with PDFs in the $\msbar$ factorization scheme. 
}
\label{tab:Z-lo-settings-val}
\end{table}

\begin{figure}[!h]
\centering
  \includegraphics[width=0.46\textwidth]{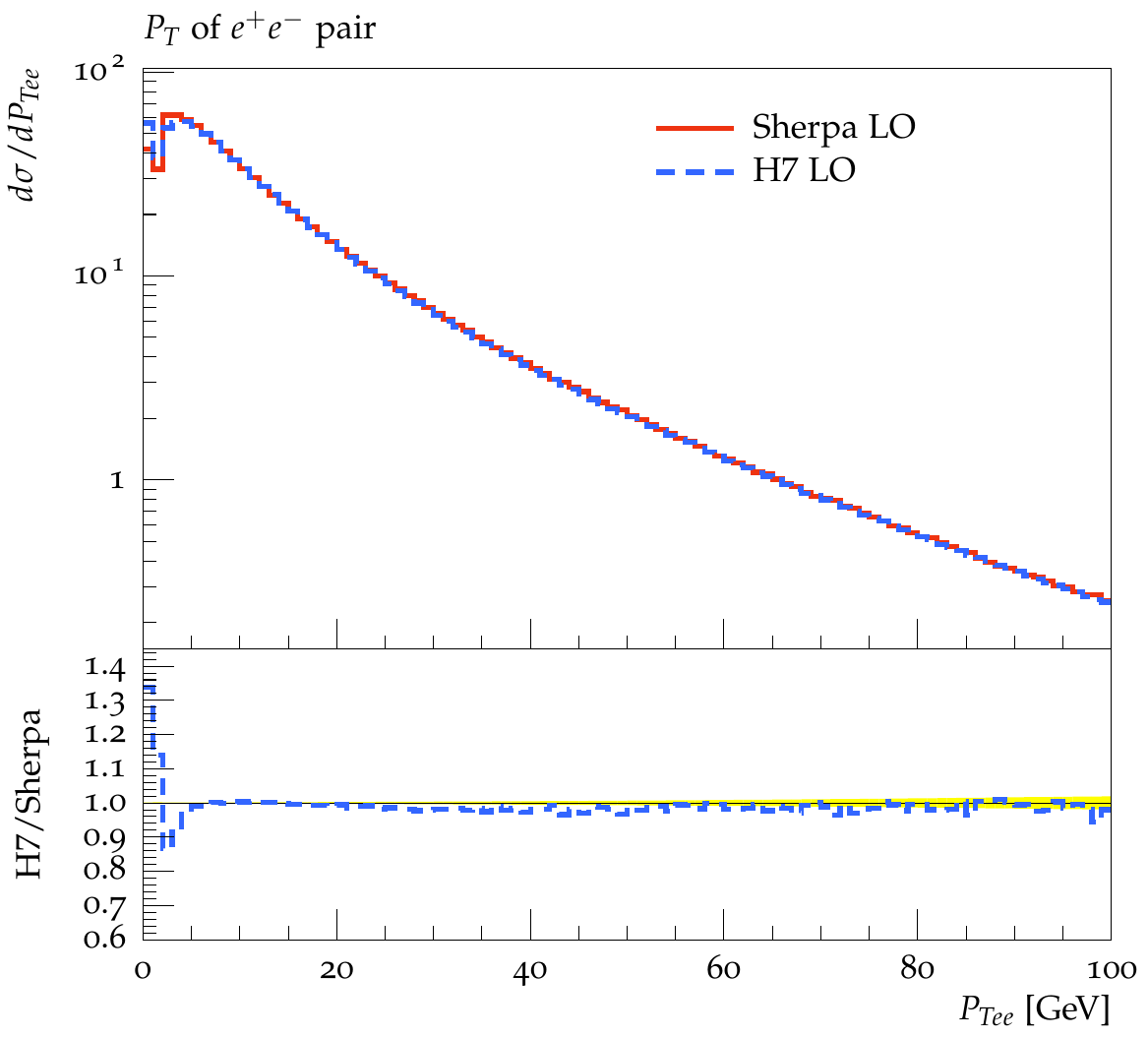}
  \includegraphics[width=0.46\textwidth]{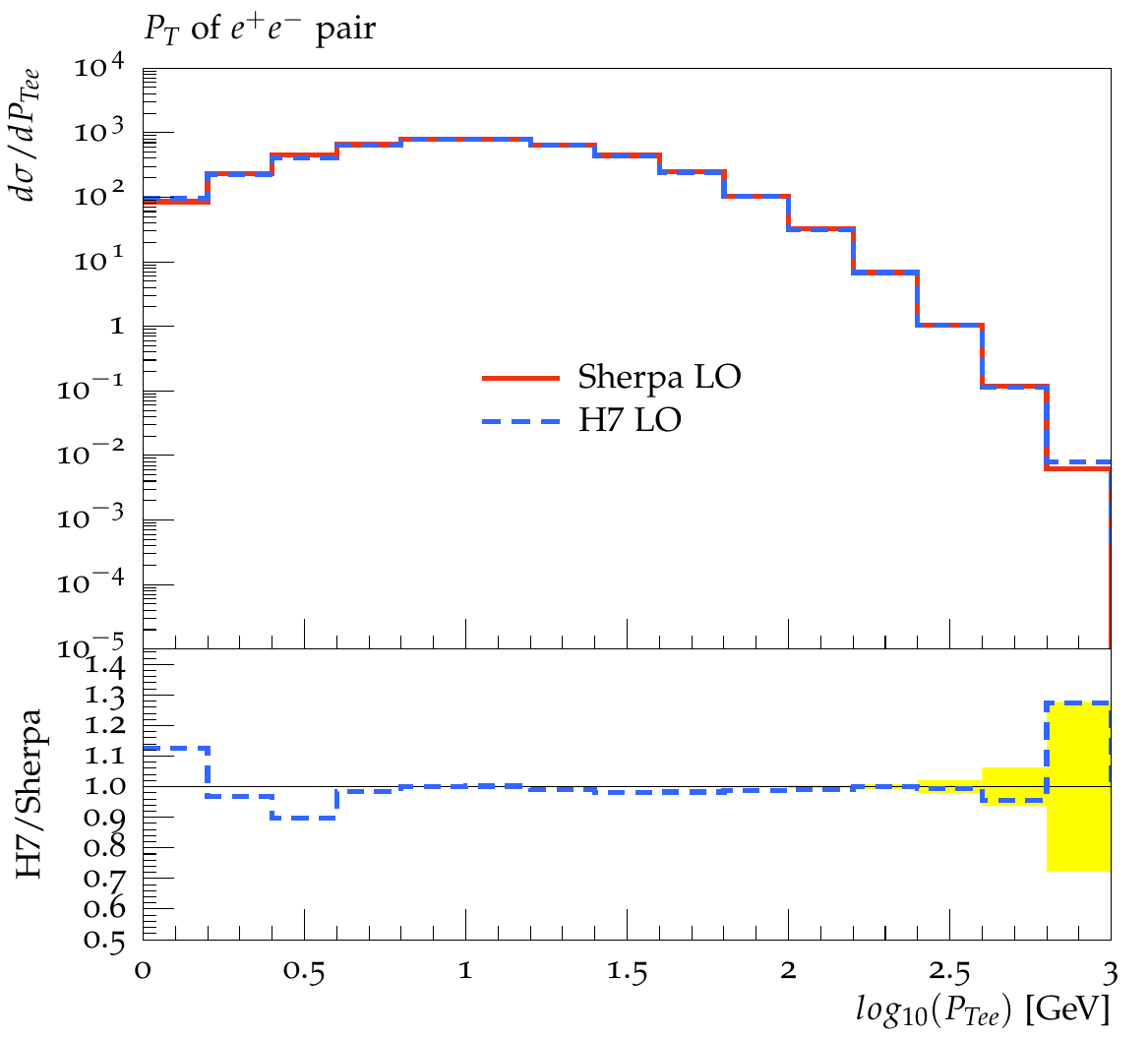}\\
   \includegraphics[width=0.46\textwidth]{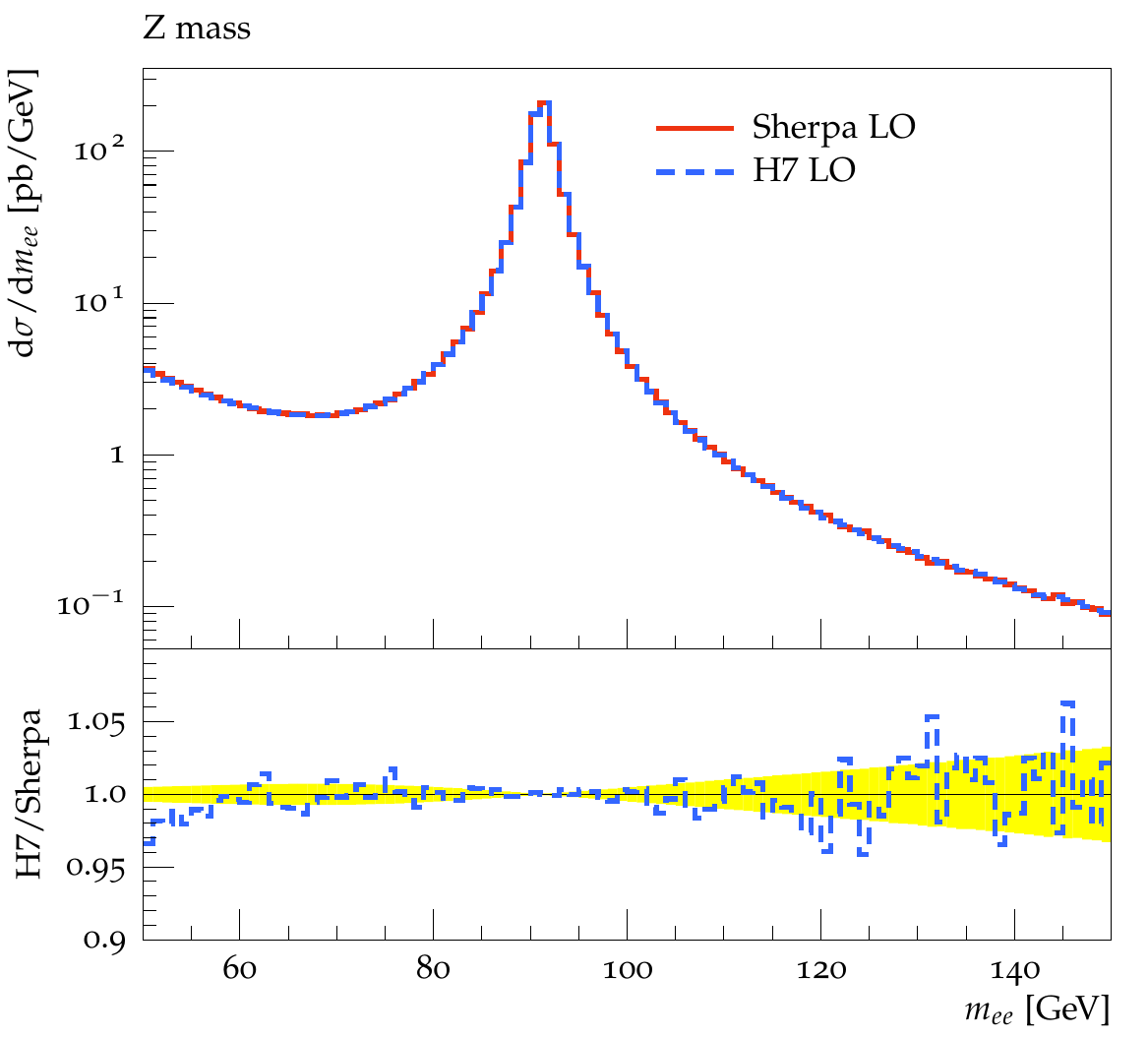}
  \includegraphics[width=0.46\textwidth]{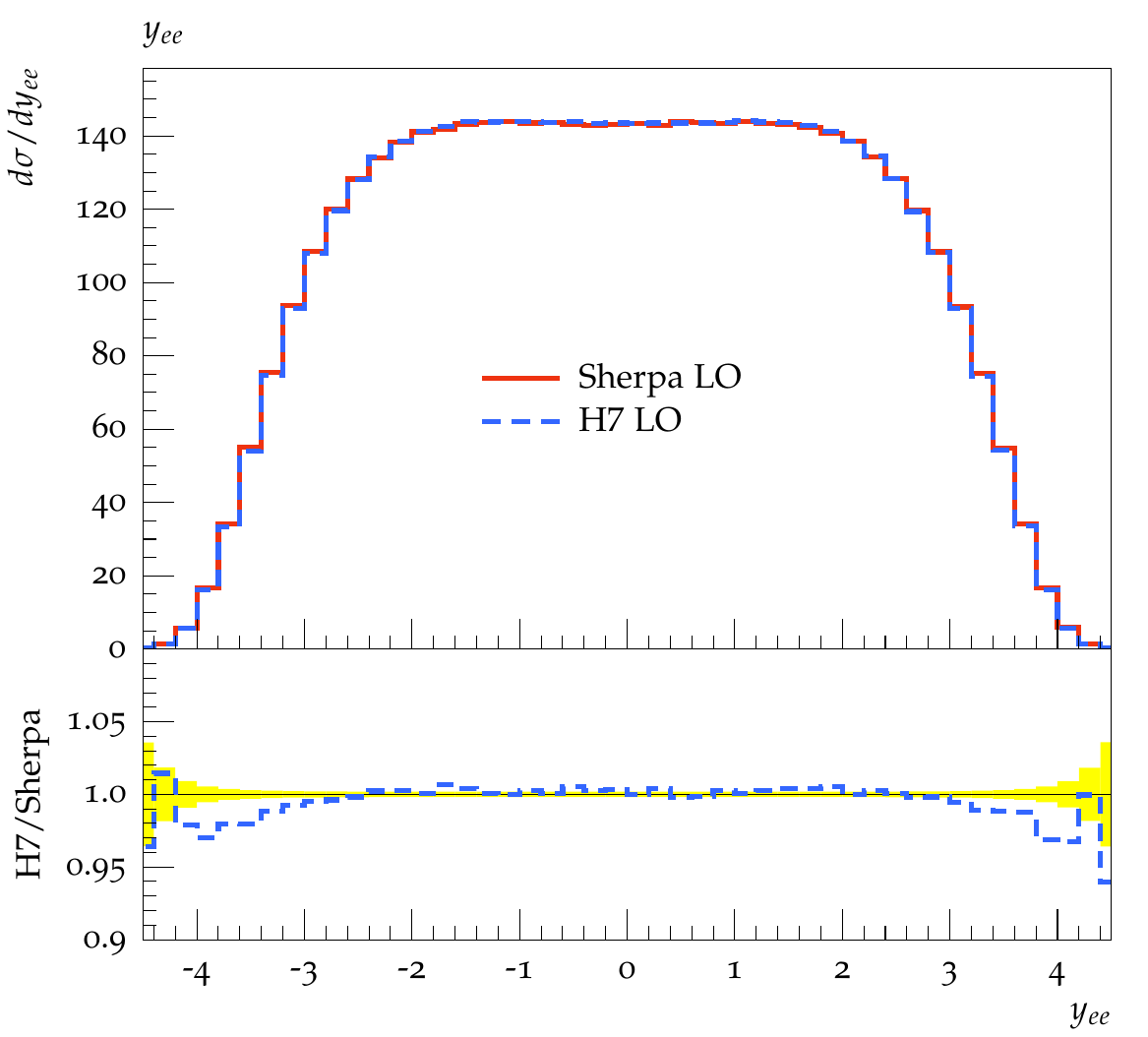}\\
\caption{\sf
Comparisons of the $Z/\gamma^*$ distributions 
from \sherpa{} and \herwig{} for the LO-level Drell--Yan process
with $e^+e^-$ pairs in the final state.
}
\label{fig:LO-Z-pt}  
\end{figure}

Since the \krknlo{} method depends on details of the parton-shower algorithm%
\footnote{For example, in \sherpa{} the initial-state parton shower has a prefactor of $1/2$ 
in the scale of the running $\as$ in the calculation 
of the Sudakov form factor.} both dipole 
showers~\cite{Schumann:2007mg,Platzer:2011bc} have to be set as  similar as possible.
The level of agreement of the two PSMCs is presented in Fig.~\ref{fig:LO-Z-pt},
where we show distributions of the final-state $e^+e^-$-pair  
(the decay product of $Z/\gamma^*$-boson) 
transverse momentum $p_{Tee}$, invariant mass $m_{ee}$ and rapidity $y_{ee}$. 
We can see that all of the distributions are
in a good agreement. In the case of the transverse momentum distribution there are
some differences at small $p_T$ that result from different treatment of intrinsic-$k_T$
and the soft-parton limit in the two programs,  where differences in the 
latter emerge from the different ordering variables employed in the two programs. 
The differences at high $p_T$ are due to limited statistics in this region.

\subsubsection{KrkNLO results: H7 vs. Sherpa implementations}
\label{sssec:dynlo}

\begin{table}[!h]
\centering
\begin{tabular}{|c||c|c|c|}
  \hline
                          & \mcfm{}  & \krknlo{} \sherpa{} & \krknlo{} \herwig{}  \\
  \hline\hline                        
  $\sigma_\text{tot}$ [pb] &  $1086.5\,(1)$     &  $1045.2\,(3)$      &  $ 1046.5\,(2)$ \\ 
  \hline
\end{tabular}
\caption{\sf
Values  of  the  total  cross  section  with  statistical  errors  
for  the  Drell--Yan  process
from  the  \krknlo{}  method implemented in \sherpa{} and \herwig{} compared  to 
the  fixed-order  result  of  \mcfm{}. 
}
\label{tab:Z-nlo-settings-val}
\end{table}

With the consistent predictions obtained at the LO level, 
we are now ready to examine the consistency between 
the \sherpa{} and \herwig{} implementations of \krknlo{}. For the comparisons, we 
consider both the $q\qbar$ and $qg$ NLO channels of the DY process, 
with the backward evolution of the parton shower  
running to the end, as in the standard PSMC set-up.
In this set-up the argument of $\as$ in the hard-real NLO corrections is
the evolution variable $q^2$, \ie $\as(q^2)$, 
and in the virtual+soft-real correction it is set to $M_Z$, \ie $\as(M_Z^2)$. 
Here both programs use the $\rm MC_{DY}$ version of PDFs, 
those in which only the parton--parton transitions
relevant to the DY process up to NLO are considered, see Ref.~\cite{Jadach:2015mza}.

Once again, we start from the comparison of the total cross section 
-- the results are collected in 
Table~\ref{tab:Z-nlo-settings-val}.
We see that both implementations of the \krknlo{} method give cross sections
that agree at the per-mille level -- the tiny residual discrepancy is due to the 
aforementioned differences in the low-$p_T$ region between \herwig{} and \sherpa{}
which affect the \krknlo{} correcting weights. 
These values also agree with our previous results in Ref.~\cite{Jadach:2015mza}
(see Table~5 there).

\begin{figure}[!h]
\centering
  \includegraphics[width=0.46\textwidth]{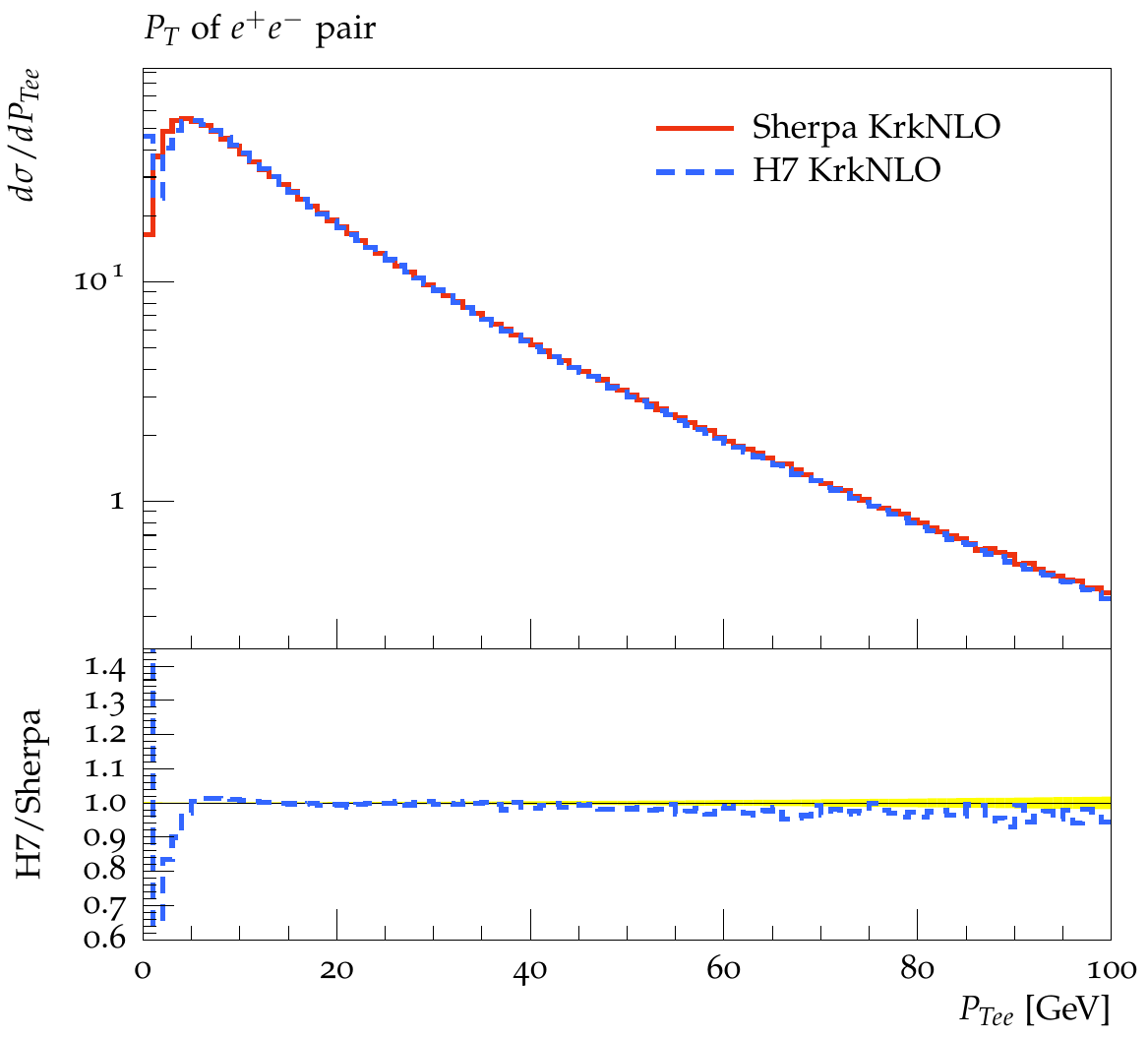}
  \includegraphics[width=0.46\textwidth]{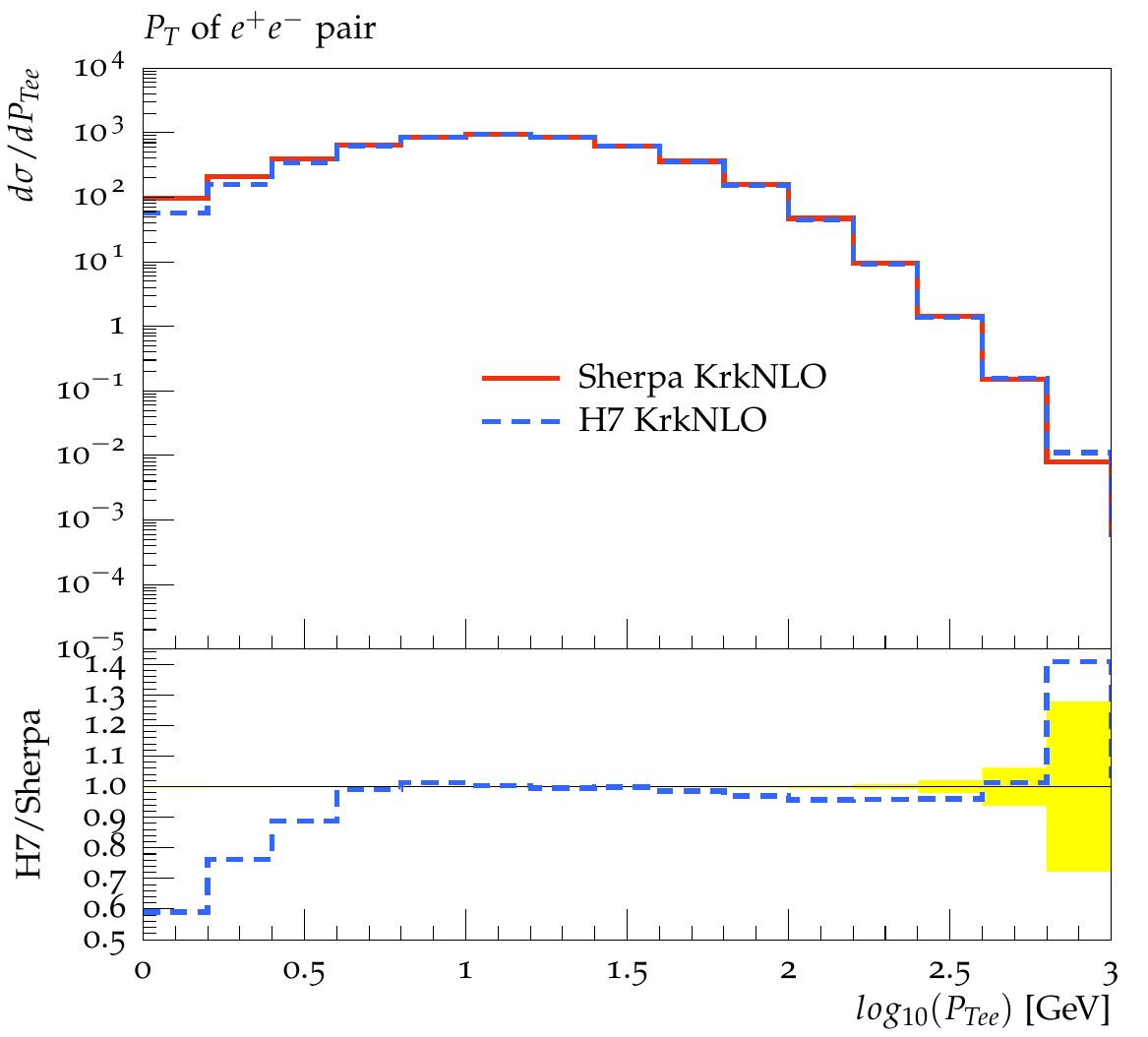}\\
   \includegraphics[width=0.46\textwidth]{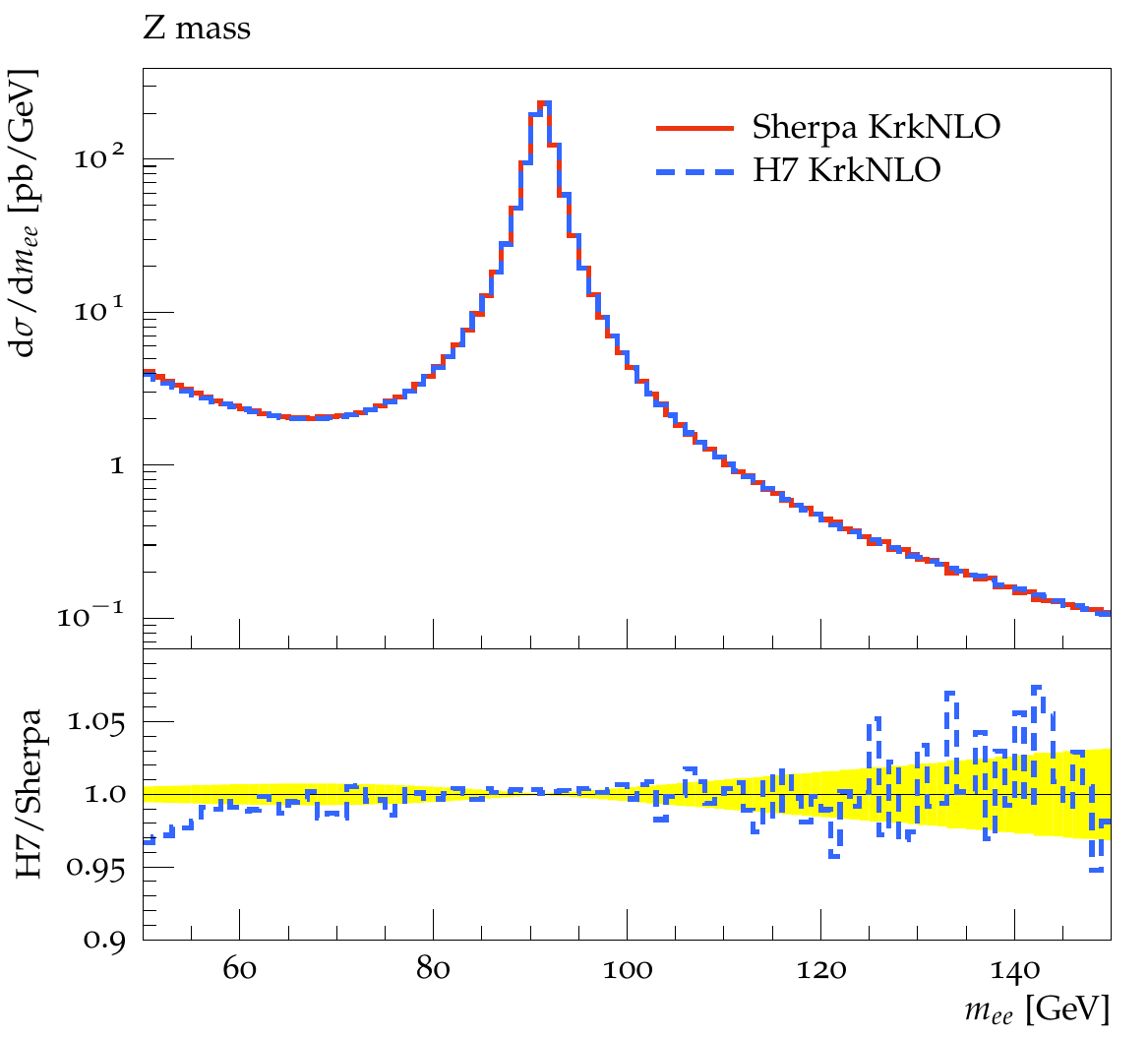}
  \includegraphics[width=0.46\textwidth]{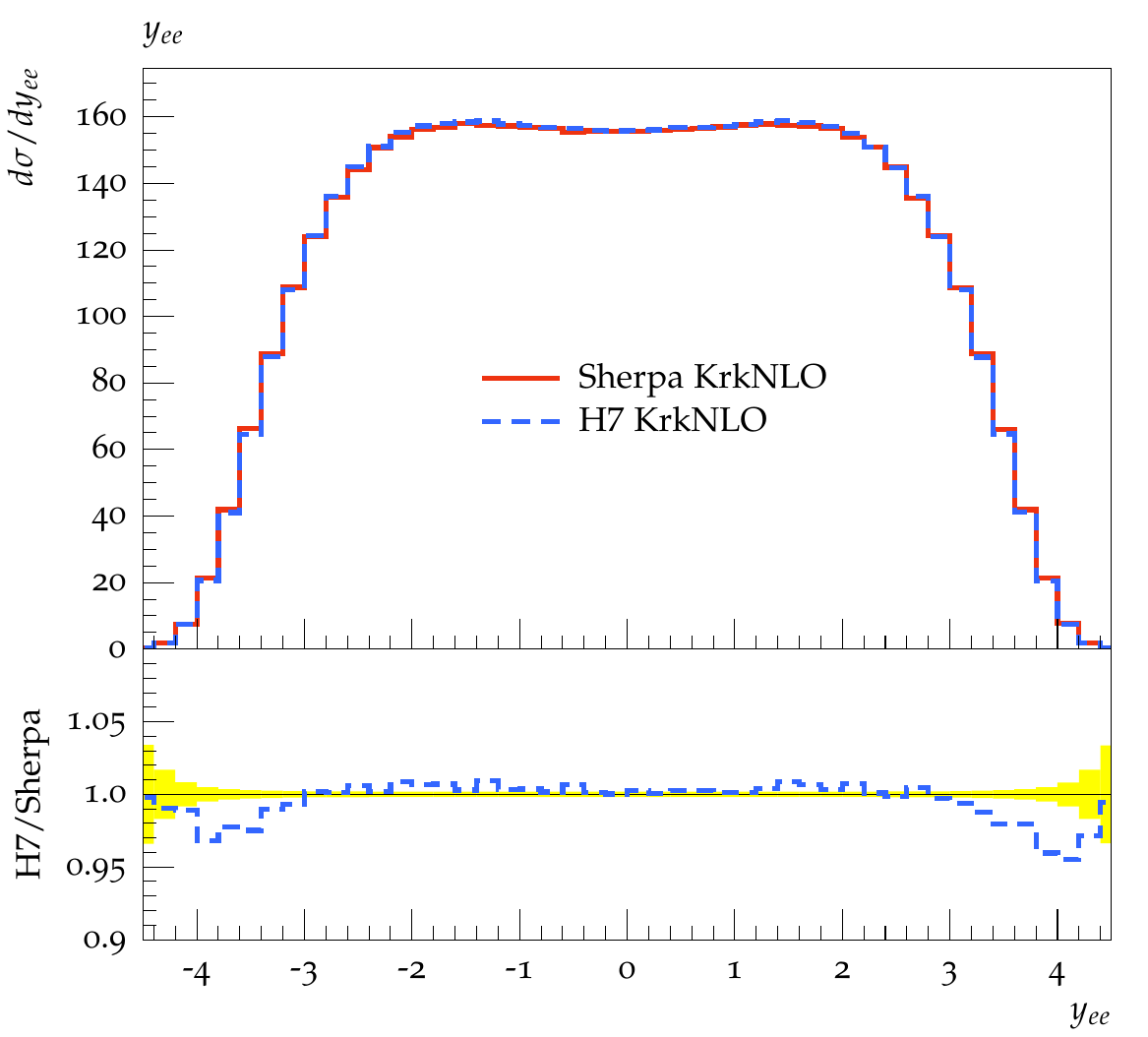}\\
\caption{\sf
Comparisons of the $Z/\gamma^*$ distributions from the
\krknlo{} method, as implemented in \sherpa{} and \herwig{}, 
for the Drell--Yan process with $e^+e^-$ pairs in the final state, 
see text for details.
}
\label{fig:NLO-2ch-pt}
\end{figure} 

In Fig.~\ref{fig:NLO-2ch-pt}, we show similar distributions 
as in Fig.~\ref{fig:LO-Z-pt}, but this time include
the NLO QCD corrections according to the \krknlo{} method. 
Again, a good agreement between the two programs is found. 
Only in the low $p_T$ region of the $p_{Tee}$ distributions are some differences visible, 
but they reflect effects already seen in Fig.~\ref{fig:LO-Z-pt}. 
Given this agreement we are able to validate our implementation 
of the \krknlo{} method in \herwig{}.

\subsubsection{KrkNLO results: MC vs. $\bf MC_{DY}$ prescriptions}
\label{sssec:dymc}

\begin{table}[!h]
\centering
\begin{tabular}{|c||c|c|c|}
  \hline
  & \mcfm{}: $\msbar$ PDFs & \krknlo{}: $\rm MC_{DY}$ PDFs & \krknlo{}: MC  PDFs\\
  \hline \hline                        
  $\sigma_\text{tot}$ [pb] &  $1086.5\,(1)$     &  $1046.5\,(2)$      
                                                  &  $1022.3\,(2) $ \\ 
  \hline
\end{tabular}
\caption{\sf
Values  of  the  total  cross  section  with  statistical  errors  
for  the  Drell--Yan  process from \krknlo{} 
implemented in \herwig{} for two variants of MC-scheme PDFs 
compared  with the  fixed-order NLO result from \mcfm{}, 
see text for details. 
}
\label{tab:Z-nlo-PDF-MC-MCDY}
\end{table}

\begin{figure}[!h]
\centering
  \includegraphics[width=0.46\textwidth]{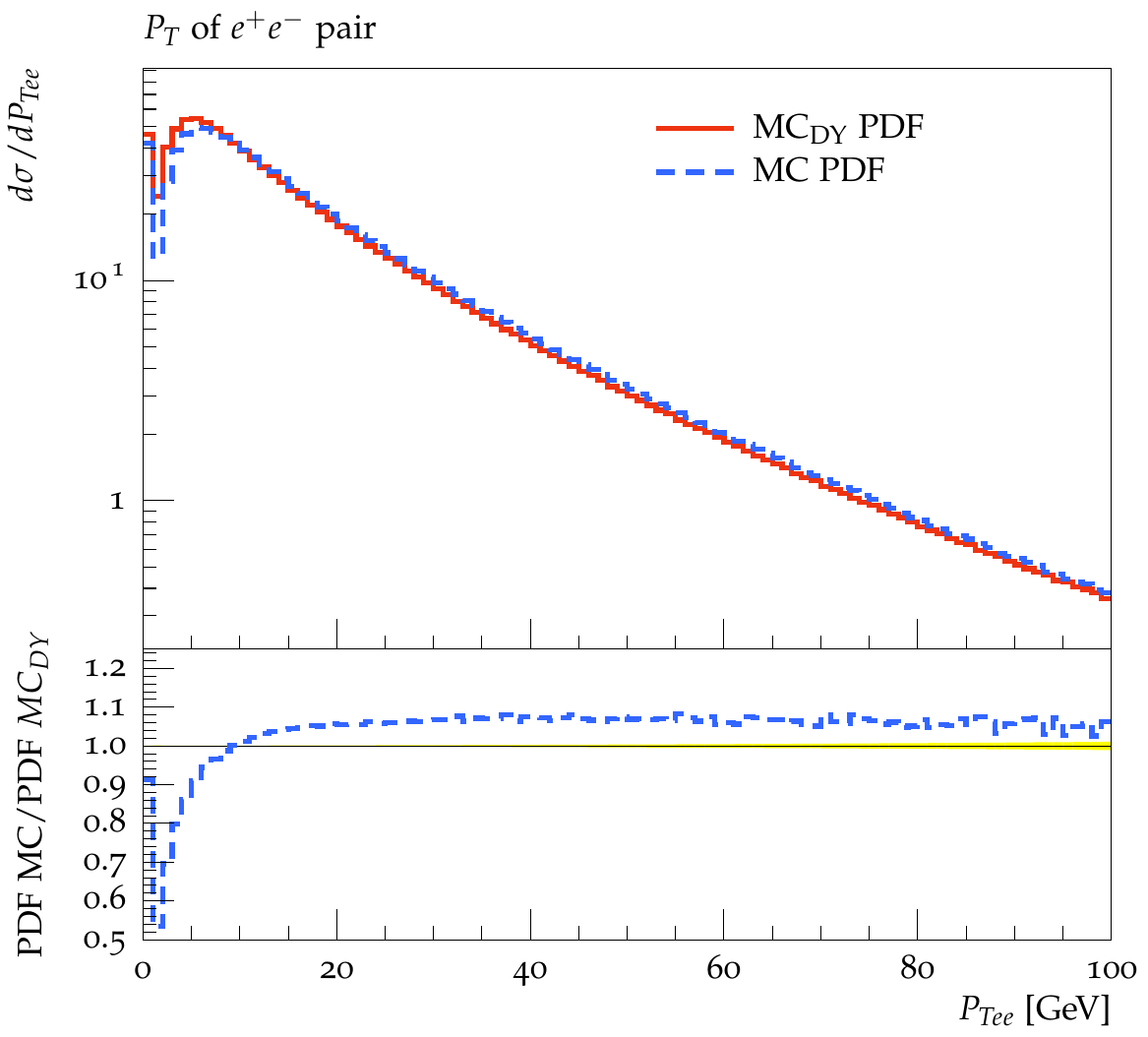}
  \includegraphics[width=0.46\textwidth]{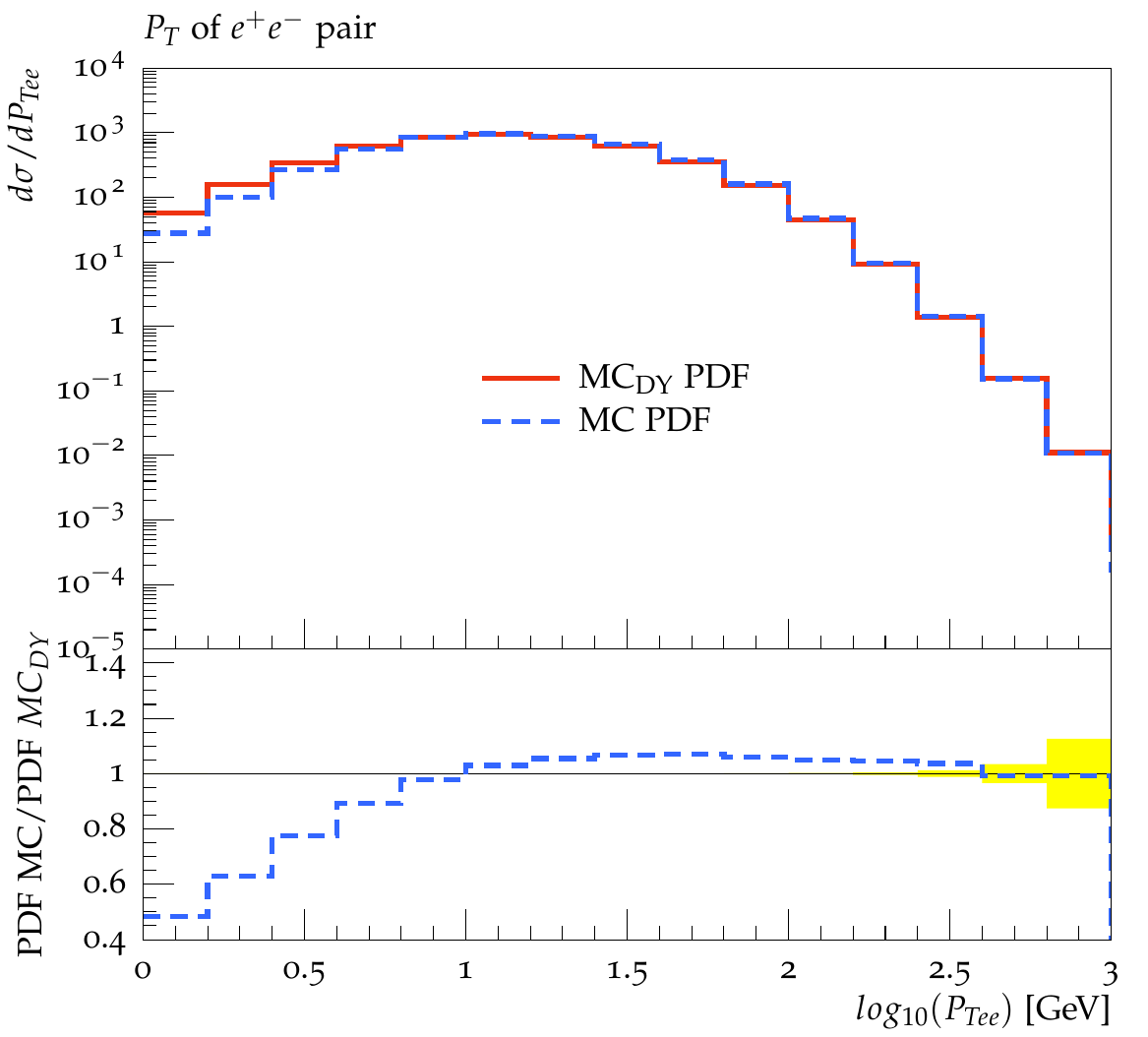}\\
   \includegraphics[width=0.46\textwidth]{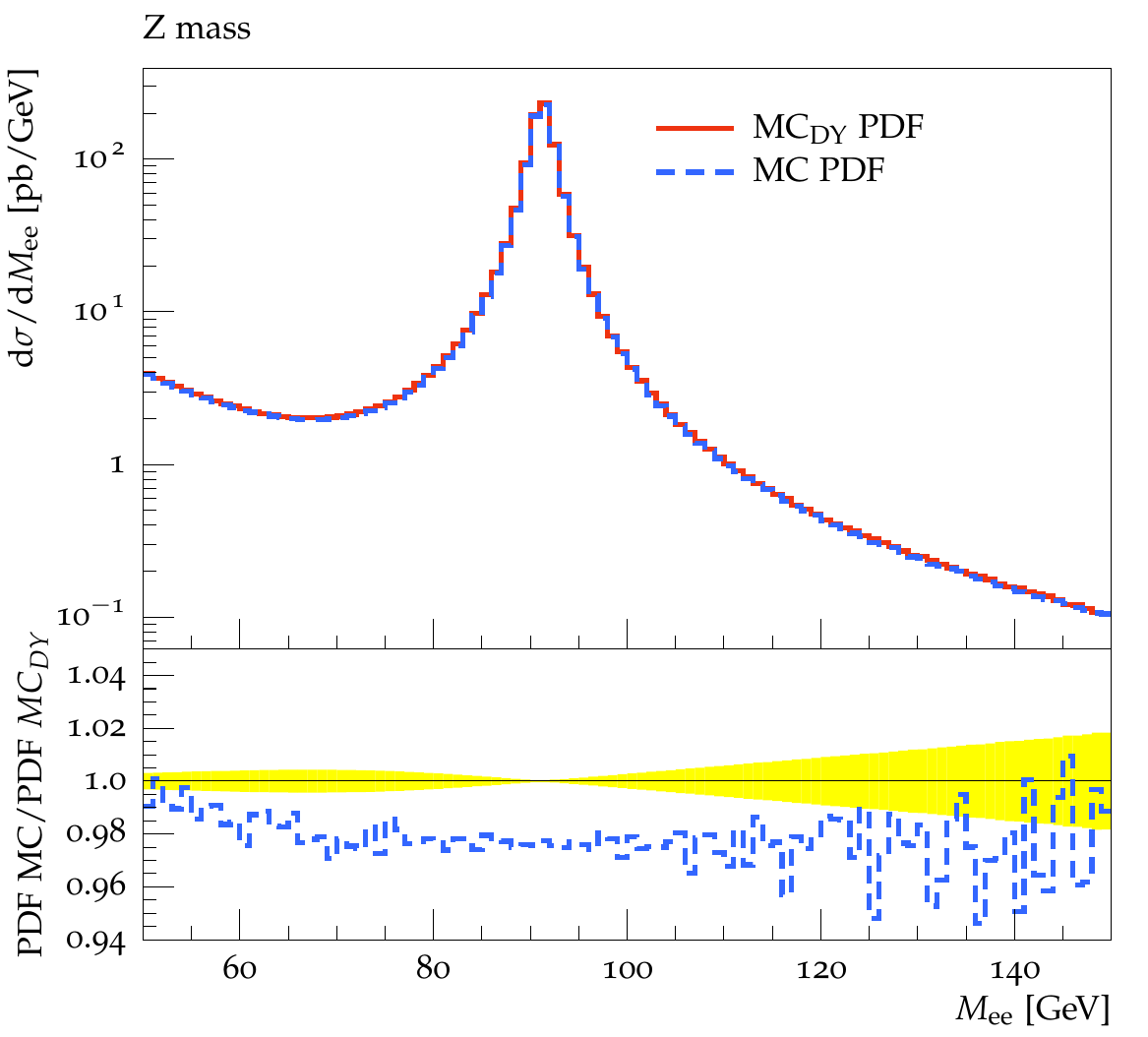}
  \includegraphics[width=0.46\textwidth]{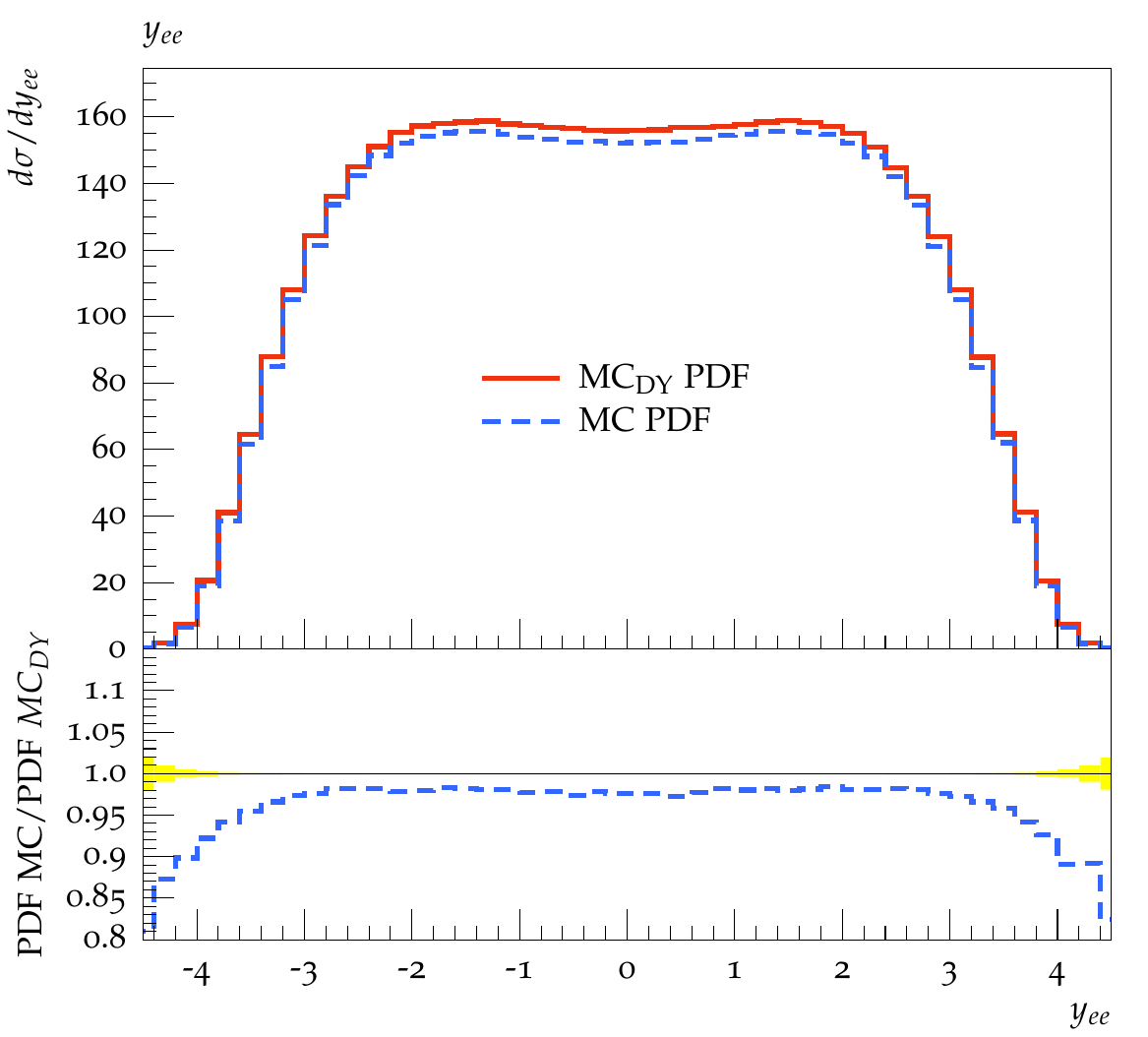}\\
\caption{\sf
Comparisons of the $Z/\gamma^*$ distributions, as in the previous plots, 
from \krknlo{} implemented in \herwig{} 
for both the $\rm MC_{DY}$ PDFs and the complete MC PDFs.
In the \krknlo{}, the NLO weights use $\as(q^2)$ 
in the hard-real corrections and $\as(M_Z^2)$ in the virtual+soft-real ones.
}
\label{fig:NLO-2ch-pt-PDF-MC-MCDY}
\end{figure} 

Having validated the implementation of the \krknlo{} method in \herwig{} we are
ready to present the first new results. We start from the comparison of the
\krknlo{} predictions based on the $\rm MC_{DY}$ PDFs, defined in
Ref.~\cite{Jadach:2015mza}, with those in which the complete MC-scheme PDFs,
first introduced in Ref.~\cite{Jadach:2016acv}, are used.  The main difference
between the MC$_\text{DY}$ and MC PDFs is that in the former 
only the quark PDFs are transformed from
the $\msbar$ to MC scheme and the gluon PDF is unchanged, whereas in the latter
the gluon PDF is also transformed to the MC scheme; 
this is described in Ref.~\cite{Jadach:2016acv}.  
 
We note that for the DY process the transformation of
the gluon PDF comes as an effect beyond NLO, so formally, for any predictions at
the NLO level, the $\rm MC_{DY}$ PDFs are sufficient.  
However, this is not the case for processes 
in which  initial-state gluons are present at the LO level, as is the case for
Higgs boson production in gluon--gluon fusion. Therefore, for future
applications of the \krknlo{} method to a generic process we shall use the
complete MC PDFs of Ref.~\cite{Jadach:2016acv}.

From Table~\ref{tab:Z-nlo-PDF-MC-MCDY} we see that the differences 
between the values of the total cross section 
corresponding to these two variants of the MC-scheme PDFs are rather small, 
$\sim 2\%$, well within uncertainties of the NLO predictions. For comparison, 
we also show the fixed-order NLO result obtained from \mcfm{} using the $\msbar$ PDFs. 
The differences with respect to the \krknlo{} results are at the level of
$4$--$6\%$, also within the NLO accuracy. 
For the distributions of the quantities as shown in the previous plots, 
presented in Fig.~\ref{fig:NLO-2ch-pt-PDF-MC-MCDY}, 
the differences are at the level of a few per-cent, 
except for in the low $p_T$ region where they can grow up to $\sim 50\%$, 
but this region is very sensitive to soft gluon effects 
(and thus to the gluon PDF) that are formally beyond the NLO approximation.

\subsection{Higgs-boson production}
\label{ssec:higgs}

In this section we present results for Higgs-boson production in
gluon--gluon fusion at the LHC obtained with the \krknlo{} method implemented in
\herwig{} and compare them with predictions of other NLO-PSMC matching
approaches, namely \mcatnlo{}~\cite{Frixione:2006he} and
\powheg{}~\cite{Frixione:2007vw} as implemented in \herwig{}, 
as well as with fixed-order calculations at
NLO and NNLO from \hnnlo{}~\cite{Catani:2007vq}
and an NNLL+NNLO calculation from \hqt{}~\cite{deFlorian:2011xf,Bozzi:2005wk}.
We opt to use the {\tt CT10nlo} PDF
set \cite{Lai:2010vv} which has $\as(M_Z^2)=0.118$, and set \herwig{}
equivalently, such that we have a consistent $\as$ setting across all
predictions\footnote{Aside from small differences in the running $\as$
between \hnnlo{} and \herwig{}.}.

Finally, we confront the theoretical predictions 
of all the above matching methods with experimental
measurements performed at the LHC during Run~1 
by the ATLAS collaboration~\cite{Aad:2015lha}.

\subsubsection{LO results}
\label{sssec:hlo}
\begin{table}[!h]
\centering
\begin{tabular}{|c||c|c|c|}
  \hline
   &      \hnnlo{@LO} & \herwig{} \\
\hline\hline
  $\sigma_\text{tot}$ [pb] &     $5.565\,(1)$    &  $ 5.564\,(1)$ \\
  \hline
\end{tabular}
\caption{\sf
Values of the total cross section with statistical errors (in brackets) at the LO level for 
Higgs production in gluon--gluon fusion at the LHC 
for the $\msbar$ CT10nlo ($\as=0.118$) PDFs from \hnnlo{} and \herwig{} (fixed order),
see text for details.
}
\label{tab:H-lo-settings-val}
\end{table}

We start from the numerical cross-check 
at the LO level of different programs used in our study.
In Table~\ref{tab:H-lo-settings-val} we show the results for the total cross section obtained
from \hnnlo{} and \herwig{} (fixed order). These values are in very good agreement,
despite small differences in the running of $\as$.%
We are therefore assured that all of the parameters 
as well as PDFs used in computation of Higgs-boson production
in gluon--gluon fusion are consistently set in these programs.

\subsubsection{NLO results}
\label{sssec:hnlo}

\begin{table}[!h]
\centering
\begin{tabular}{|c||c|c|c|c|c|c|}
  \hline
   &   & \multicolumn{2}{c|}{\powheg} & &  \multicolumn{2}{c|}{\hnnlo{}} \\ 
         \cline{3-4} \cline{6-7}
   & \mcatnlo{}  & Default & Original &\krknlo{}  & NLO & NNLO \\
  \hline \hline
  $\sigma_\text{tot}$ [pb] &  $ 12.700\,(2)$ &  $ 12.699\,(2)$ 
          & $12.697\,(2) $ &  $12.939\,(2) $    & $12.640\,(1)$  & $17.063\,(15)$\\
  \hline
\end{tabular}
\caption{\sf
Values of the NLO total cross section with statistical errors (in brackets) for Higgs-boson production in gluon--gluon fusion at the LHC for \krknlo{}, \mcatnlo{} and \powheg{} as calculated by \herwig{},
as well as the NLO and NNLO result from \hnnlo{}, see text for details.
}
\label{tab:H-nlo-settings-val-cteq}
\end{table}
Here we present the results from \krknlo{} alongside those of the \mcatnlo{} and 
\powheg{} methods implemented in \herwig{} as well as the NLO and NNLO results from
\hnnlo{} as well as a results from \hqt{}. 
The \krknlo{} setup uses $\as(q^2)$ for the hard-real corrections and $\as(M_H^2)$ for the virtual+soft-real corrections.
We show two variants of the \powheg{}
method: The first one, \powheg{~(Default)},  is the default set-up in \herwig{}
and it restricts the transverse momentum of parton-shower emissions to be less than
the factorization scale, as is done in the \mcatnlo{} setup, which follows the 
initial work of Ref.~\cite{Frixione:2002ik};  the second one, \powheg{~(Original)}, 
is closer to its original implementation~\cite{Frixione:2007vw}  which has no such 
restriction. 
This amounts to \powheg{~(Default)} generating both S and H-events, with 
\powheg{~(Original)} only generating S-events.
%

\begin{figure}[h]
\centering
  \includegraphics[width=1.0\textwidth]{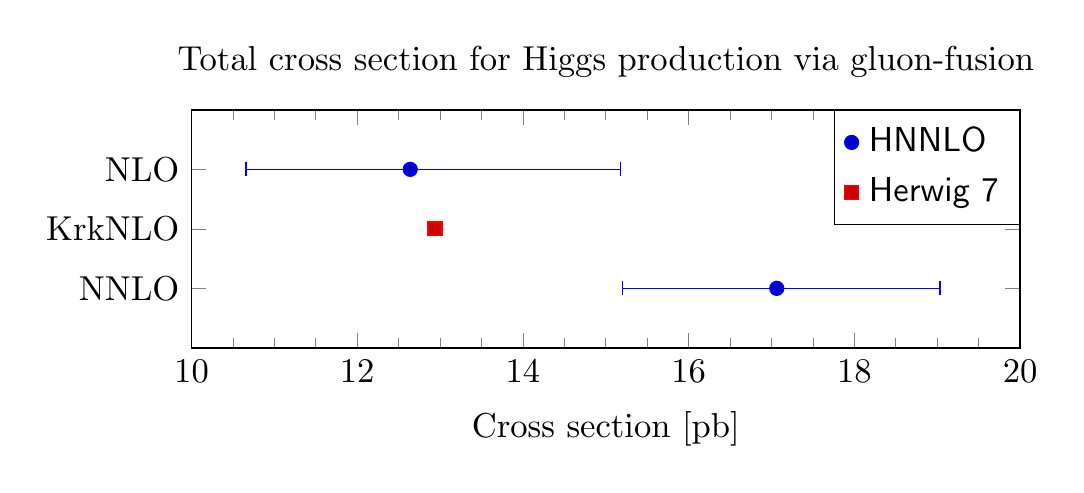}
\caption{\sf
Comparison of the total cross-section for Higgs-boson production in gluon-gluon
fusion at NLO, from \hnnlo{} and \krknlo{}, as well as at NNLO from \hnnlo{}. The
error bars, shown for \hnnlo{}, are obtained from the independent variations of
the renormalization and factorization scales by factors of $1/2$ and $2$ from $M_H$.
}
\label{fig:NLO-H-XS}
\end{figure}
\begin{figure}[h]
\centering
  \includegraphics[width=0.46\textwidth]{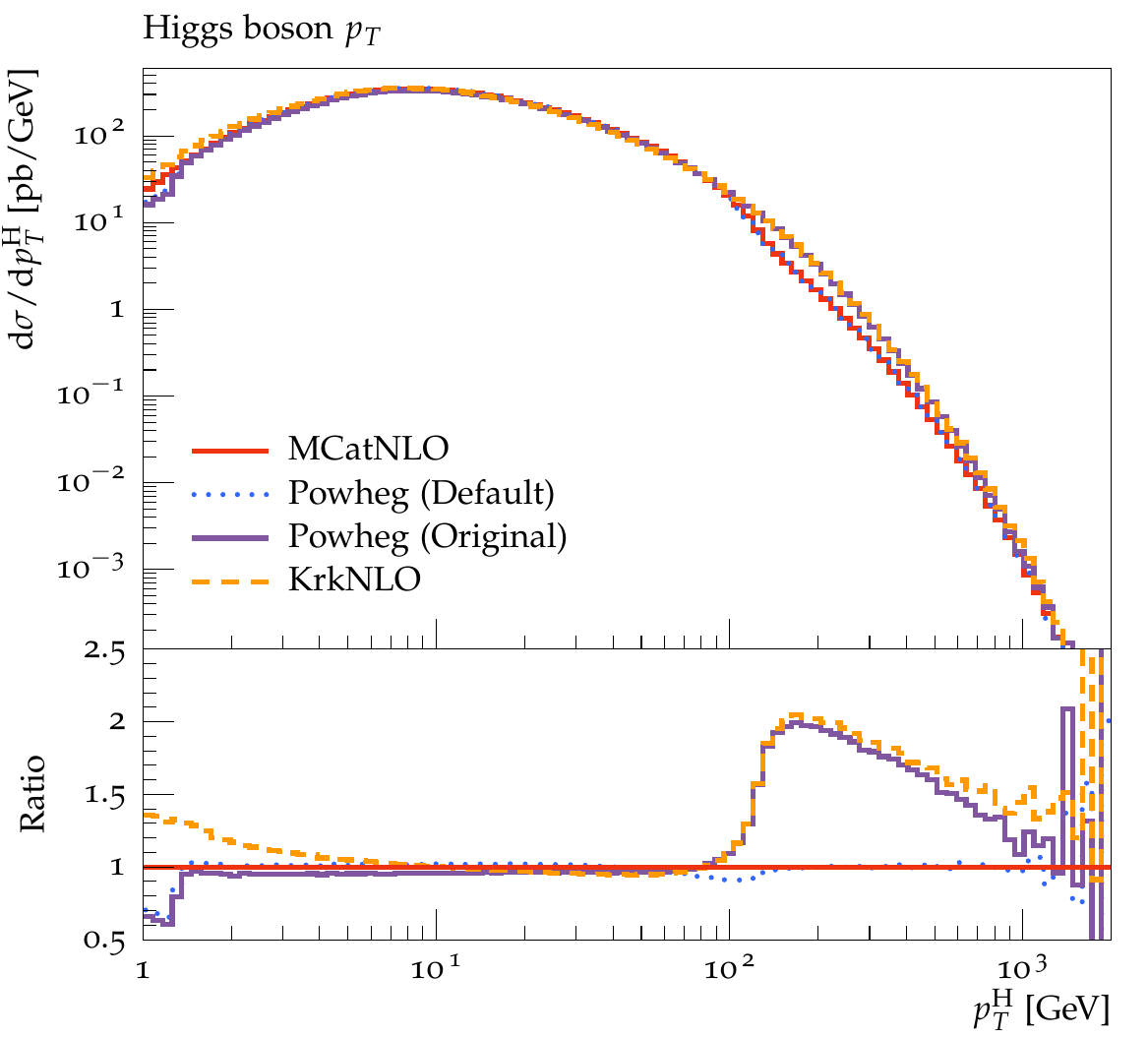}
  \includegraphics[width=0.46\textwidth]{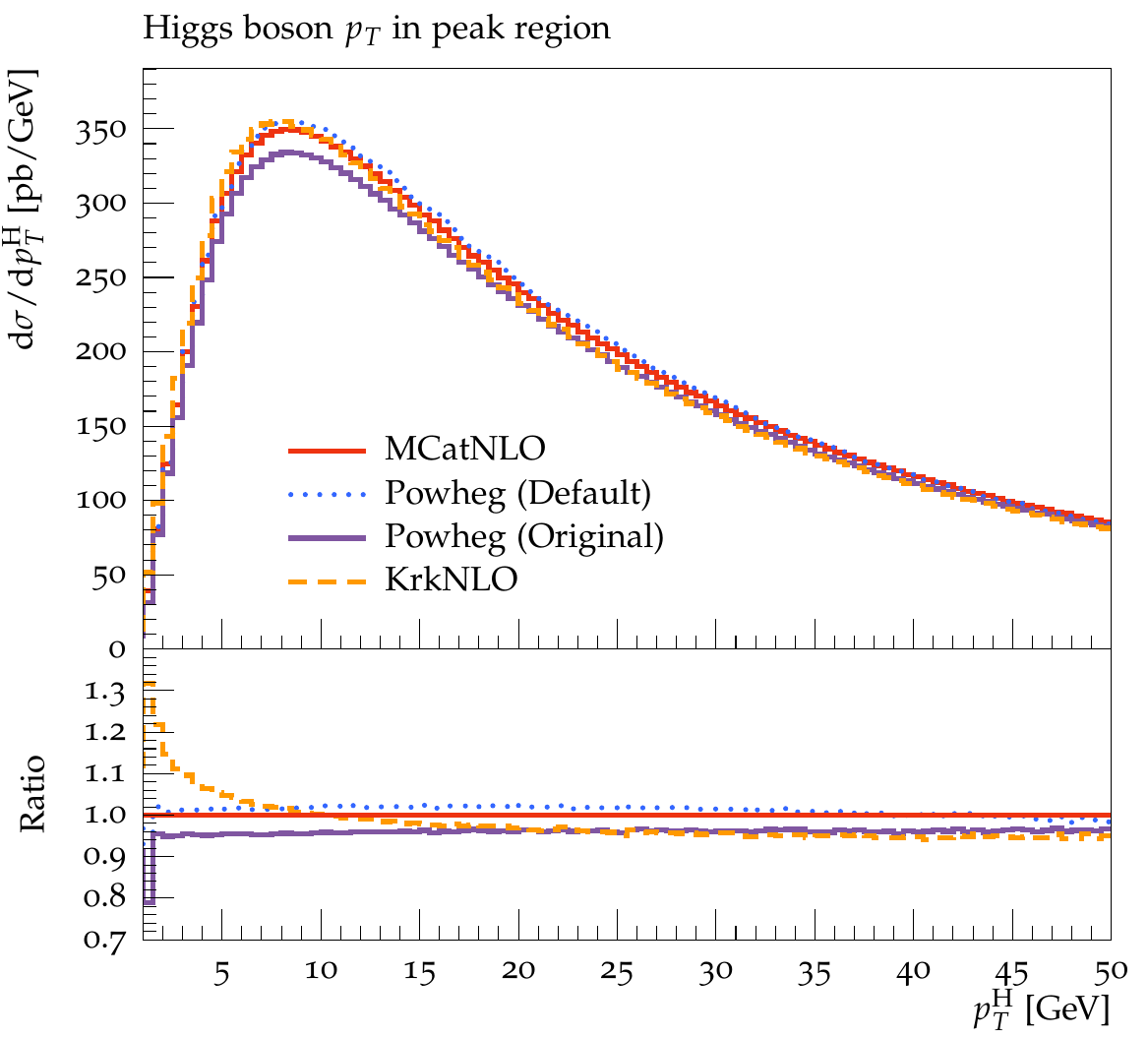}\\
  \includegraphics[width=0.46\textwidth]{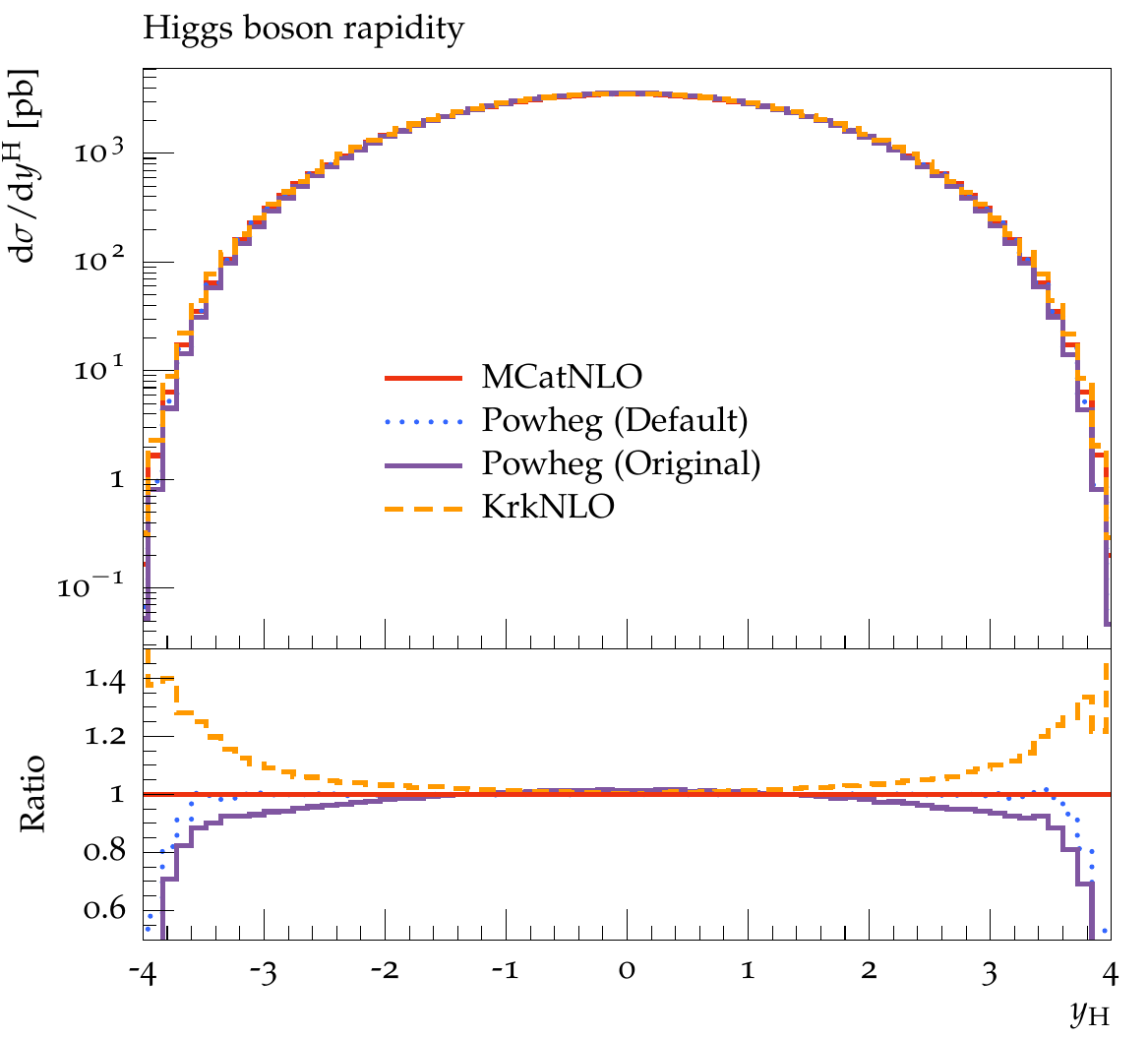}
  \includegraphics[width=0.46\textwidth]{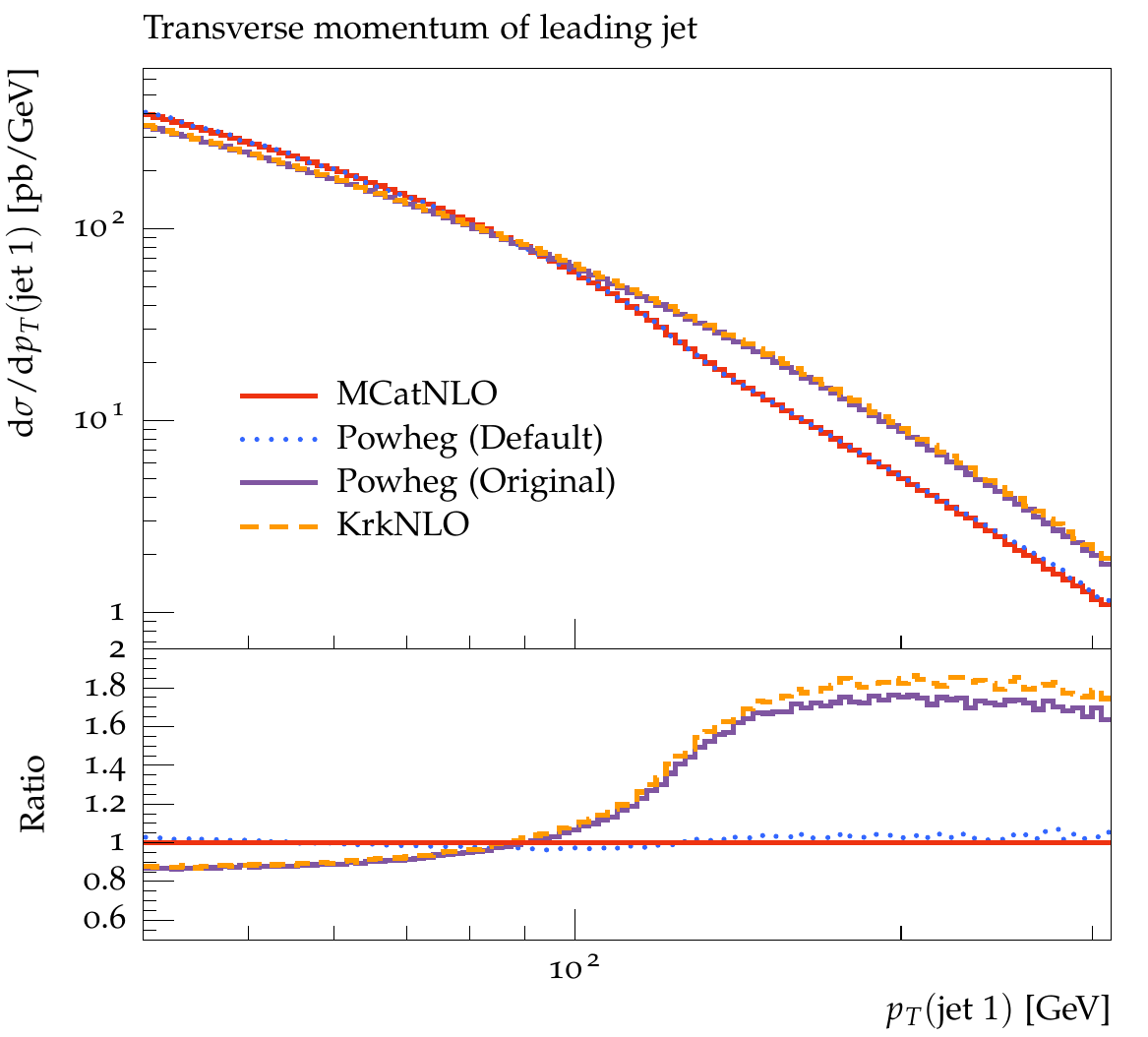}
\caption{\sf
Comparisons of the Higgs-boson transverse momentum and rapidity distributions from the
\krknlo{}, \mcatnlo{} and \powheg{} methods implemented in \herwig{} 
for Higgs-boson production in gluon--gluon fusion at the LHC, see text for details.
}
\label{fig:NLO-H}
\end{figure}

The values of the total cross section, with statistical errors, for the Higgs-boson 
production process are shown in Table~\ref{tab:H-nlo-settings-val-cteq}. 
We can see that, as expected, both the \mcatnlo{} and \powheg{} results
give the same total cross sections. 
The \krknlo{} method gives a slightly higher value of the cross section 
than the other methods. 
This can be explained by the beyond-NLO contributions that are partially
accounted for in the \krknlo{} result.
Additionally, in Fig.~\ref{fig:NLO-H-XS} we show the total 
cross-sections from \hnnlo{}, along with error bands generated by the 
variation of the renormalization and factorization scales by factors of $1/2$ and $2$
around $M_H$.
We see that the prediction of \krknlo{} is within the NLO 
uncertainty estimate. Furthermore, we see that the uncertainty estimate for NNLO 
is still rather large, at around $10\%$, and does not overlap with the NLO range.

\begin{figure}[!h]
\centering
 \includegraphics[width=0.79\textwidth]{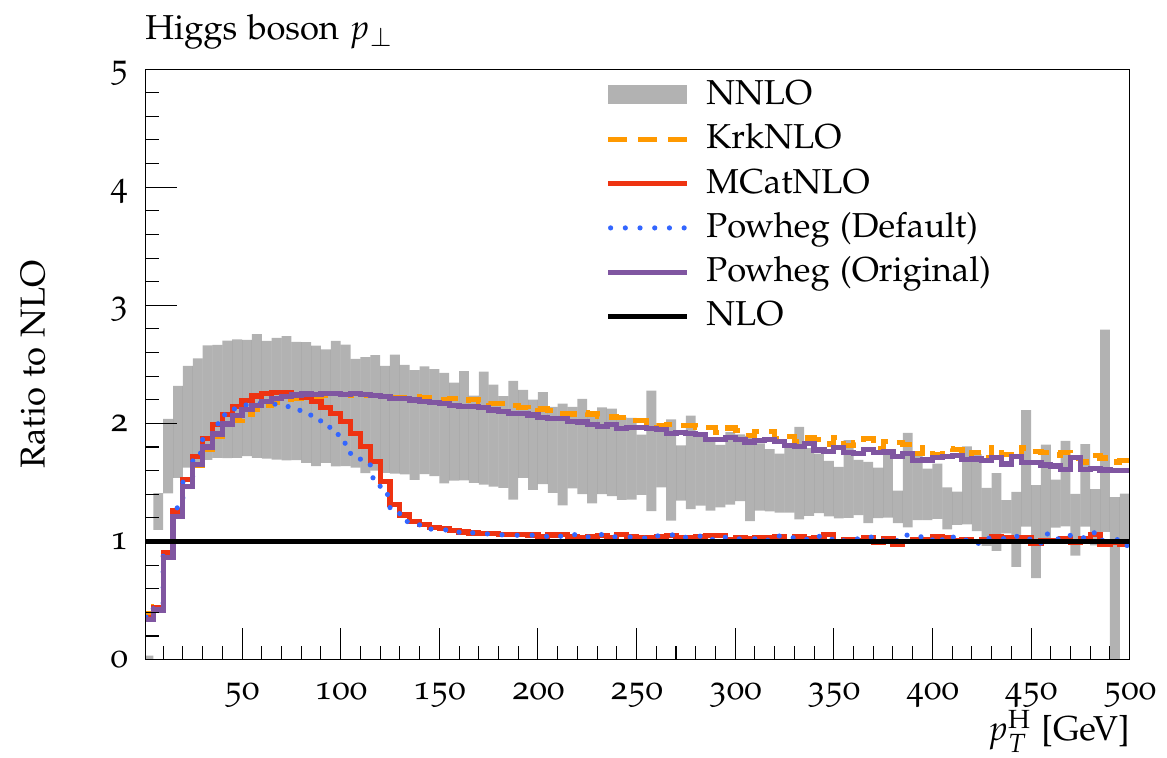} 
\\
 \includegraphics[width=0.79\textwidth]{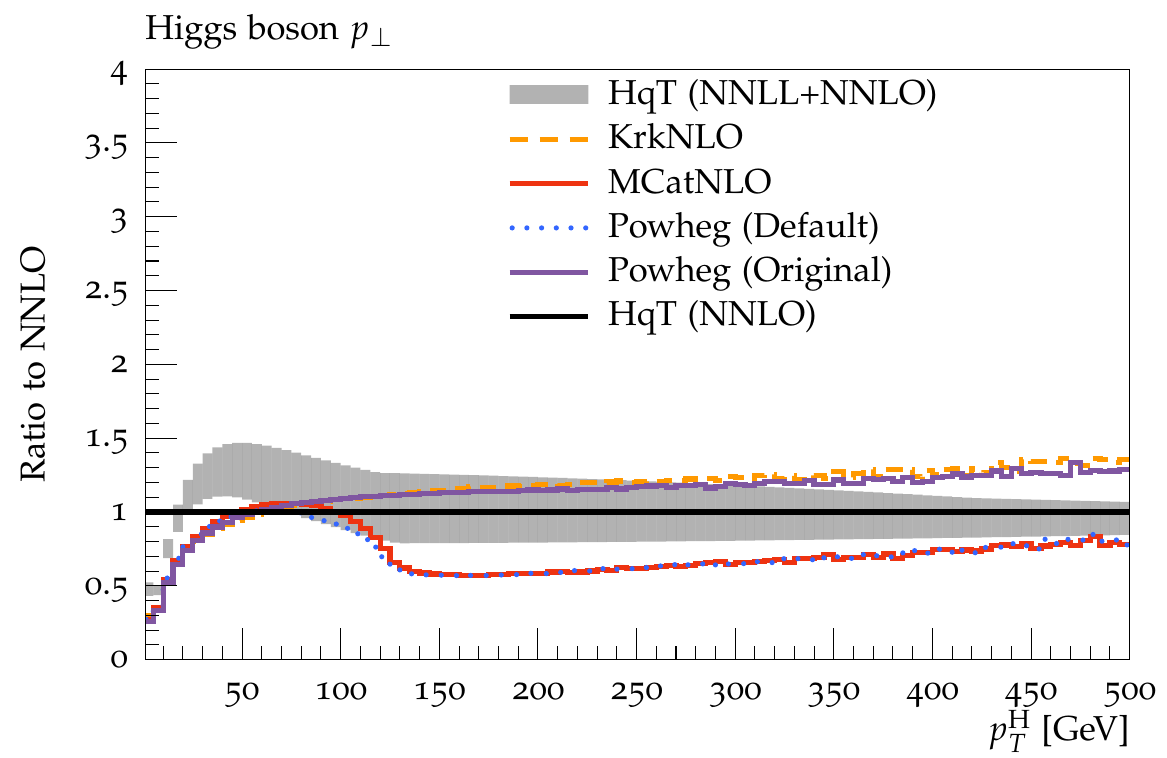}
\caption{\sf
Higgs-boson transverse-momentum distributions from
\krknlo{}, \powheg{} and \mcatnlo{}. The upper plot compares our results with the 
fixed-order NNLO result from the \hnnlo{} program and are shown relative to the 
NLO results from \hnnlo{}. The lower plot shows our results in comparison to \hqt{}, 
these are shown relative to the \hqt{} NNLO prediction. The content of the error
bands is described in the main text.
}
\label{fig:NNLO-H}
\end{figure}
\begin{figure}[t]
\centering
  \includegraphics[width=0.46\textwidth]{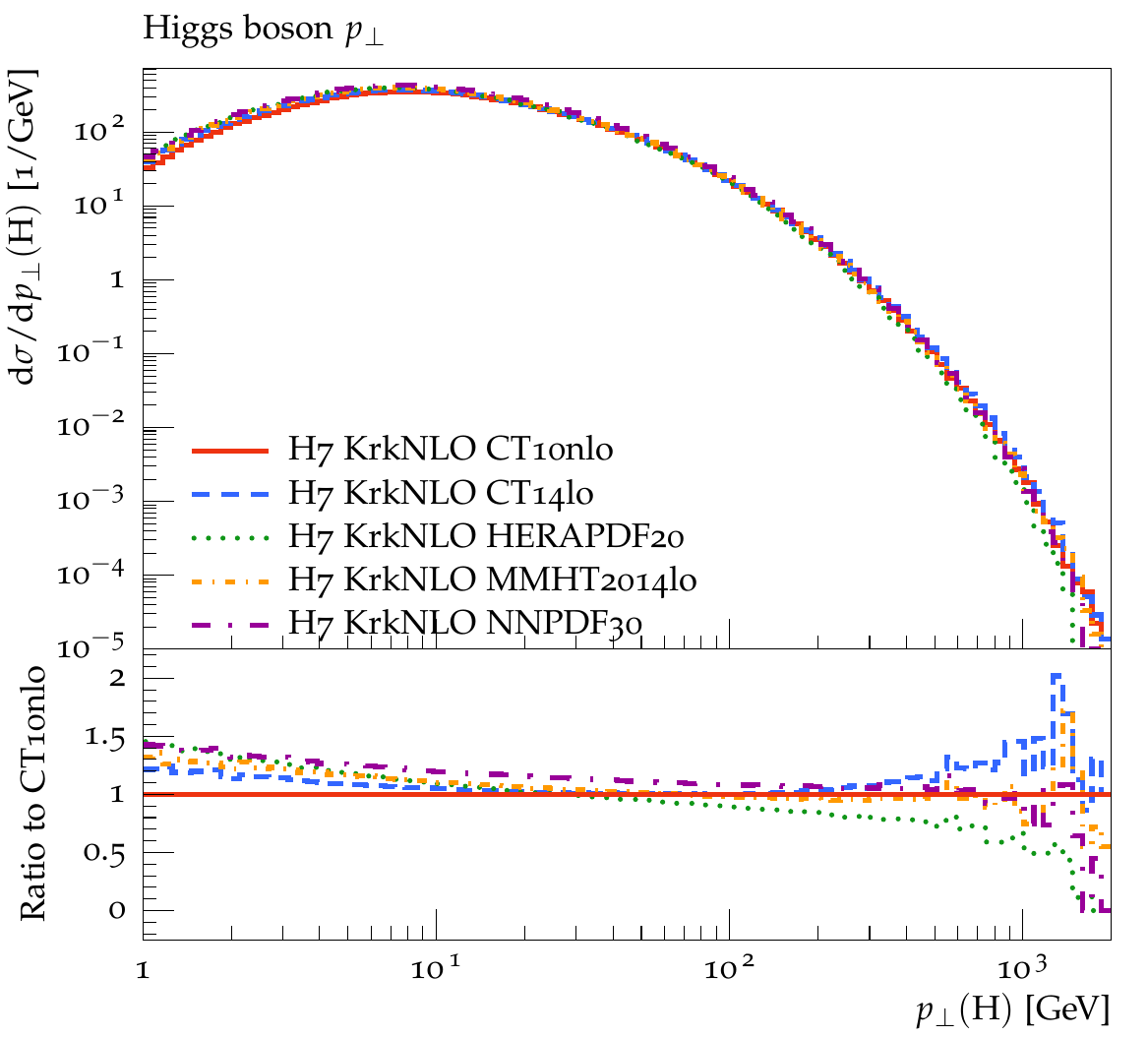}
  \includegraphics[width=0.46\textwidth]{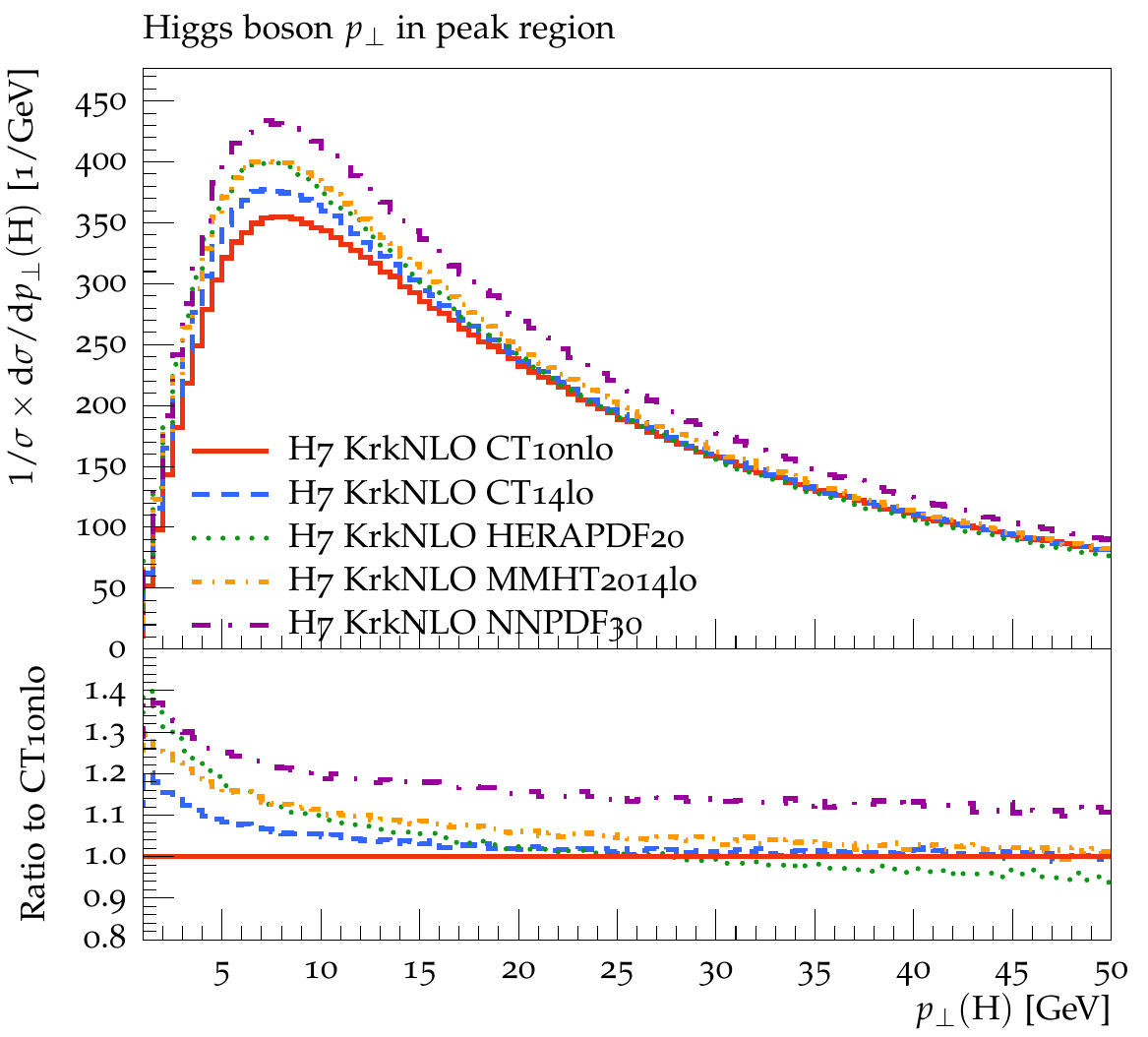}\\
   \includegraphics[width=0.46\textwidth]{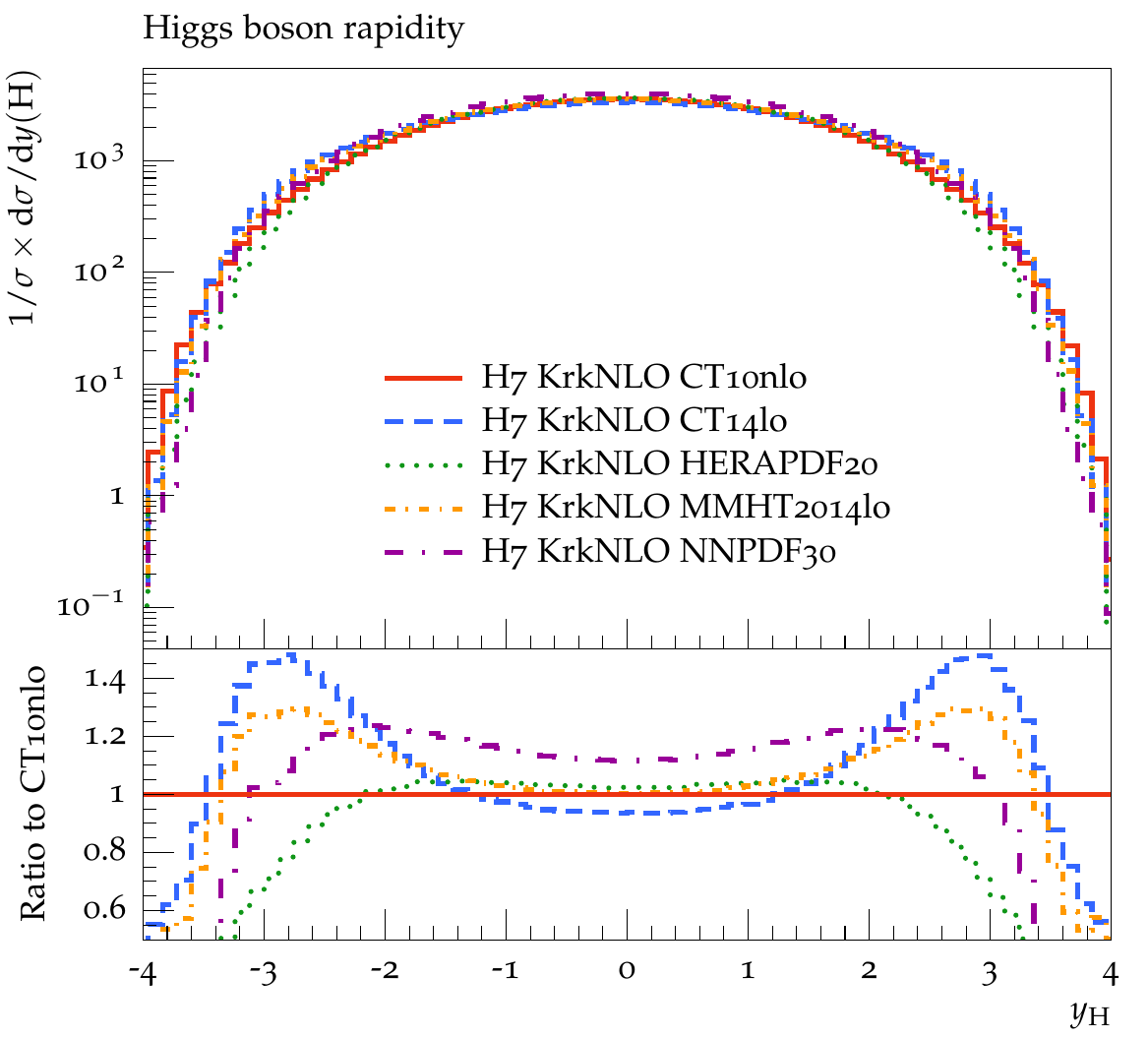}
\caption{\sf
Comparisons of the Higgs-boson transverse momentum and rapidity distributions from the
\krknlo{} method using different PDF sets in the MC factorization scheme
for the Higgs-boson production in gluon--gluon fusion at the LHC, see text for details.
}
\label{fig:NLO-H-pdf}
\end{figure}

\begin{figure}[t]
\centering
  \includegraphics[width=0.49\textwidth]{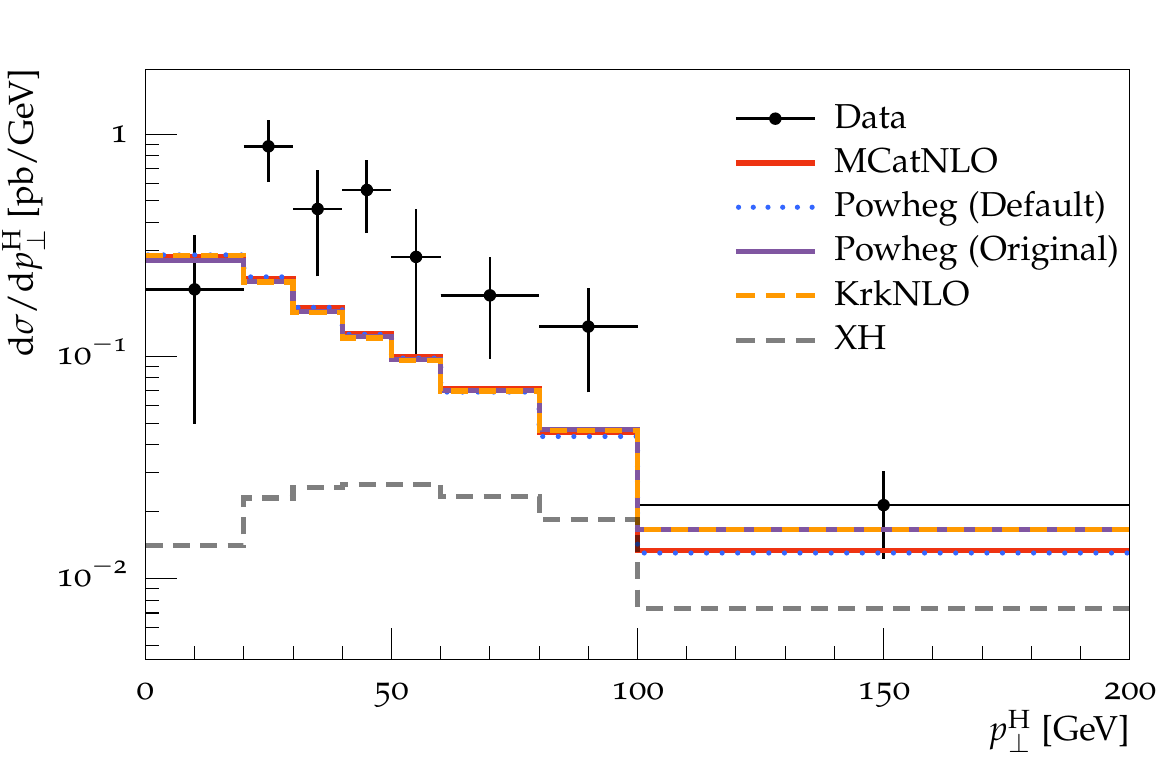}
  \includegraphics[width=0.49\textwidth]{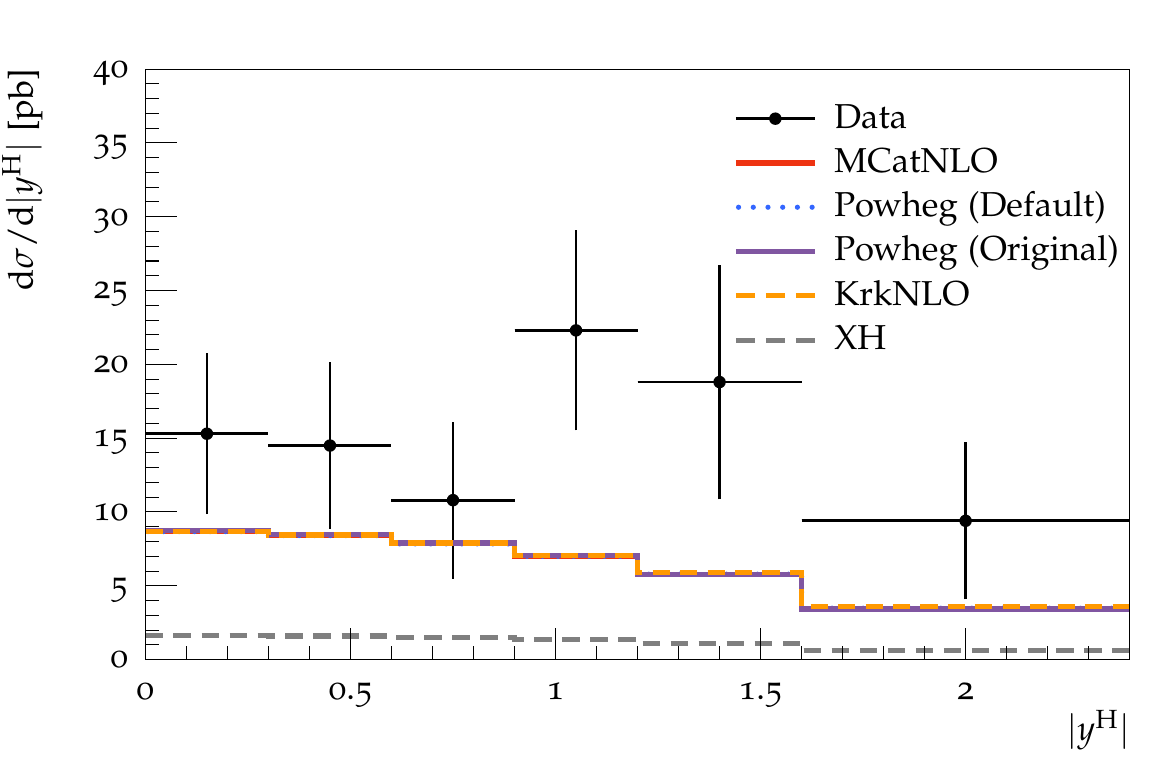}\\
   \includegraphics[width=0.49\textwidth]{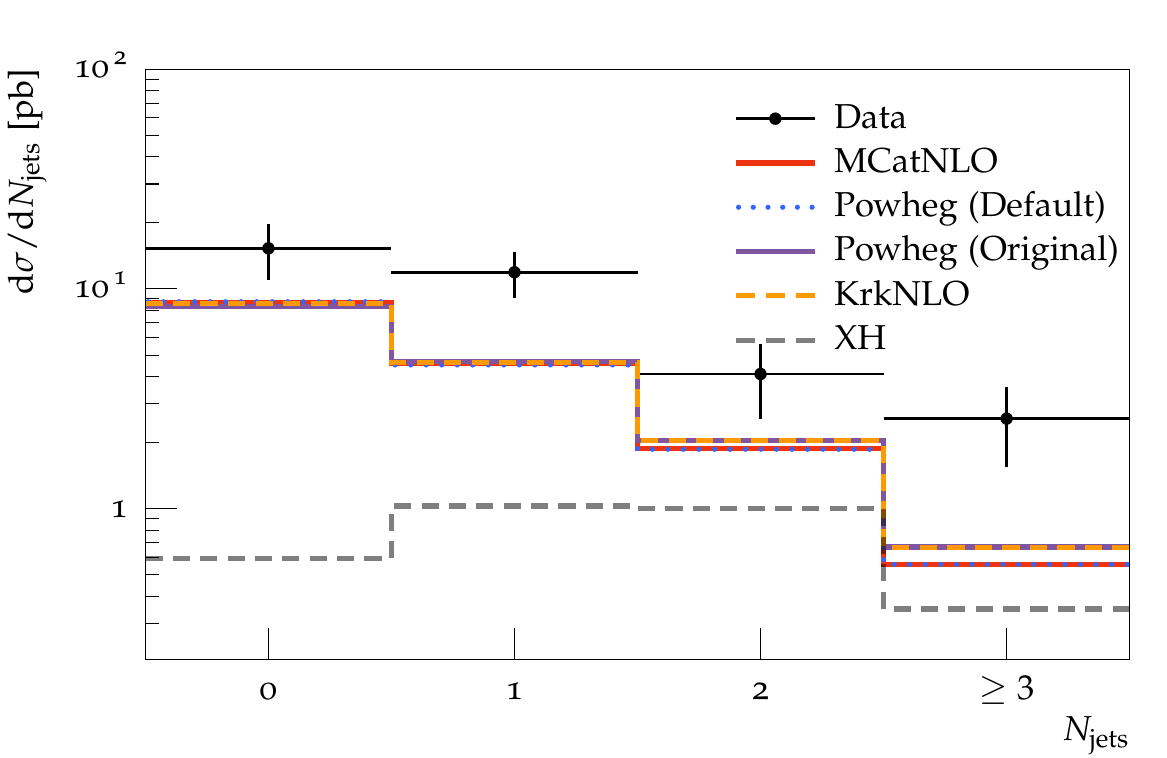}
  \includegraphics[width=0.49\textwidth]{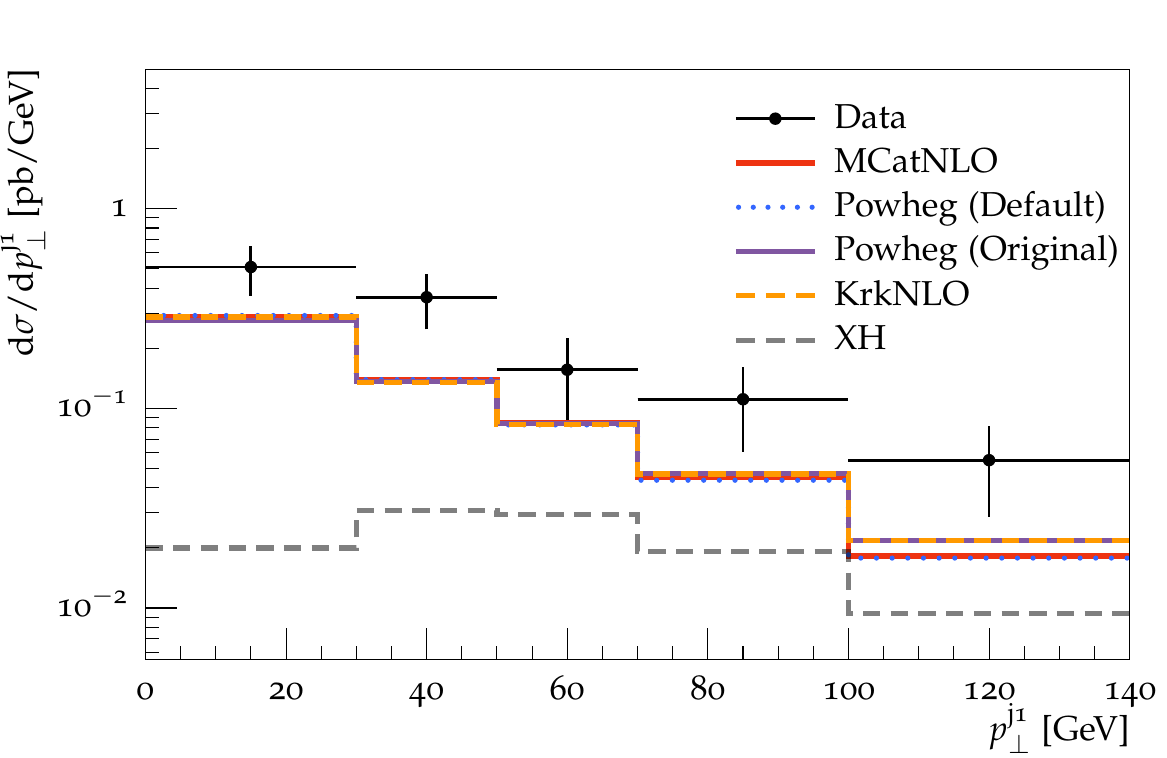}\\
\caption{\sf
Comparisons of the predictions of the \krknlo{}, \mcatnlo{} and \powheg{} 
methods implemented in \herwig{} for Higgs-boson production 
with the ATLAS data from LHC Run~1. 
The gluon--gluon fusion results from \herwig{} are plotted on top of the XH
 results from Ref.~\cite{Aad:2015lha}.
}
\label{fig:NLO-H-ATLAS}
\end{figure}

In Fig.~\ref{fig:NLO-H} we present the distributions 
of the Higgs-boson transverse momentum $p_T^\text{H}$ 
and rapidity $y_\text{H}$, comparing the results from \krknlo{} 
with the ones from \mcatnlo{} and \powheg{}.
All of the predictions agree within a few per-cent for the $p_T^\text{H}$ range from 
$\sim 5\,$GeV to  $\sim 100\,$GeV and for the rapidity range $[-3,3]$. 
In the region $p_T^\text{H} > M_H$ larger differences between \krknlo{} and 
\mcatnlo{}/\powheg{~(Default)} are visible, whereas \krknlo{} produces a similar shape to \powheg{~(Original)}.
The main reason for this is that both \mcatnlo{} and \powheg{~(Default)} restrict 
the value of the transverse momentum of parton-shower emissions 
to be below the value of $M_H$, 
whereas for \krknlo{}, and also \powheg{~(Original)}, there is no such a restriction; 
this can be seen in the upper-left plot, where a sharp drop of the $p_T$
spectrum for $p_T \gtrsim M_H$
is visible in the case where the emissions are restricted
by this hard-cutoff. However, this spectrum can be smoothed by relaxing this
condition, as shown in Refs.~\cite{Alwall:2014hca, Bellm:2016rhh}. 
Of course, such differences are acceptable within the NLO approximation.

Next, in Fig.~\ref{fig:NNLO-H} we compare \krknlo{}, 
\mcatnlo{} and \powheg{} predictions from \herwig{} for the 
Higgs-boson transverse momentum distribution 
with the corresponding result obtained from
\hnnlo{}~\cite{Catani:2007vq, Grazzini:2008tf} and \hqt{}~\cite{deFlorian:2011xf,Bozzi:2005wk}.
The error bands \hnnlo{} NNLO distributions were obtained by varying the 
renormalization and factorization scales by the typical factors of $1/2$ and $2$ 
around $M_H$ as an estimate of the uncertainty from neglected higher orders. The 
\hqt{} distributions were obtained similarly, but also include variations of the 
resummation scale of $1/2$ and $2$ at the central value of $M_H/2$ as 
recommended in Ref.~\cite{Bozzi:2005wk}.
The \hnnlo{} comparison, upper plot, is shown relative to the NLO distribution 
from \hnnlo{}  and the \hqt{} comparison, lower plot, is shown relative to the NNLO 
distribution from \hqt{}.

As we see in the upper plot of Fig.~\ref{fig:NNLO-H}, both the \krknlo{} and the 
NNLO results show the same trends, quickly raising above the NLO result at low and moderate 
$p_T^\text{H}$ and remaining above it at high $p_T^\text{H}$.
The fact that our method gives a result that is higher than the NLO one 
at high $p_T^\text{H}$
is a consequence of the mixed real-virtual $\order{\as^2}$ terms, which
constitute part of the NNLO correction and arise because of the multiplicative
nature of the \krknlo{} approach.

In the upper plot of Fig.~\ref{fig:NNLO-H} we also show similar comparisons with NNLO
for \mcatnlo{} and two versions of \powheg{}. 
The behaviour at low $p_T^\text{H}$ is close to that
from \krknlo{}. At high $p_T^\text{H}$, however, 
\mcatnlo{} and \powheg{~(Default)}, by
construction, converge to the NLO results, 
departing from the NNLO predictions. On the other hand,
\krknlo{} and \powheg{~(Original)} are closer to the NNLO predictions but for larger $p_T^\text{H}$ values 
         they are marginally harder.
The lower plot of Fig.~\ref{fig:NNLO-H} shows results from \mcatnlo{}, \powheg{} and \krknlo{} compared to the resummed calculation from \hqt{} (for the ``switched'' option).
All of the NLO+PS give similar results up to roughly 80~GeV. The \hqt{} result peaks towards lower values of $p_T^\text{H}$ compared to the other predictions.

In Fig.~\ref{fig:NLO-H-pdf} we show the results of the \krknlo{} method 
obtained for different modern $\msbar$ 
PDF sets: {\tt CT10nlo} (as used in this section), {\tt CT14lo}, {\tt HERAPDF20}, 
{\tt MMHT2014lo} and {\tt NNPDF30lo}.
Except for {\tt CT10}, we have used the LO versions of the corresponding PDF parametrizations,
since at NLO they become negative at some regions of the $x$ variable and this poses a problem for 
the \herwig{} PSMC generator.
We can see that the distributions can vary even by up to 40\%. The biggest
differences are observed at low transverse momenta and large rapidities.
In Appendix~\ref{app:pdfs} we compare all the different PDFs in the $\msbar$ and
MC schemes and show that the differences at the level of parton distribution
functions, see Fig.~\ref{fig:pdfs-comp}, are commensurate to those observed in
Fig.~\ref{fig:NLO-H-pdf} for the differential cross sections. Further studies of
systematic effects due to PDFs are left for the future.

Finally, in Fig.~\ref{fig:NLO-H-ATLAS} we compare the predictions 
for the distributions of the Higgs-boson transverse 
momentum and rapidity, the number of jets and the transverse momentum 
of the hardest jet from \krknlo{}, \mcatnlo{} 
and \powheg{} with the ATLAS data collected in LHC Run~1 
with a collision energy of $\sqrt{s} = 8\,$TeV~\cite{Aad:2015lha}.
To our generated results for the gluon--gluon fusion we have added the contributions
from other Higgs-production channels, denoted XH, taken from Ref.~\cite{Aad:2015lha} 
-- they account for $\sim 12\%$ of the cross section.
All of the data points lie above the theoretical predictions, 
although the experimental errors are large, 
The NLO-PSMC matching methods offer largely equivalent predictions
with \krknlo{} and \powheg{~(Original)}  
predicting slightly harder spectrum for high $p_\perp$ and higher rates for larger jet multiplicities
(similar trends are also seen in Fig.~\ref{fig:NLO-H}).
Further differences were previously
discussed in the context of Fig.~\ref{fig:NLO-H}.
In our simulations we have used the {\tt CT10nlo} PDF parametrization, the same that was used 
in Ref.~\cite{Aad:2015lha} for theoretical predictions. However, we have checked that
changes of our results due to the use of different PDF sets, discussed in Appendix A,
are much smaller than the experimental errors of the ATLAS data and negligible
compared to the differences between these results and the data points.

\section{Conclusions and outlook}

In this paper we have presented the numerical results 
of the \krknlo{} method for the Drell--Yan (DY) process and Higgs-boson
production in gluon--gluon fusion at the LHC 
for the collision energy  $\sqrt{s} = 8\,$TeV.

Firstly, we have validated the implementation of the \krknlo{} method 
in the \herwig{} PSMC by comparing its results 
for the DY process with previous results obtained with the \sherpa{}-based implementation.
Then, we have presented new results for the DY process 
with the complete MC-scheme PDFs that were recently defined
in Ref.~\cite{Jadach:2016acv}. 
These results have been compared with those for the older variant of the MC-scheme PDFs,
called here the $\rm MC_{DY}$ PDFs, 
that were introduced in Ref.~\cite{Jadach:2015mza} exclusively for the DY process. 
The agreement between the results for these two variants of the MC-scheme PDFs 
has been found at the level of $\sim 2\%$, which is well within the NLO accuracy.

Our main numerical results in this paper concern the Higgs-boson production process. 
We have presented the results for the total cross section, 
as well as distributions of the Higgs-boson transverse momentum
and rapidity. 
The \krknlo{} predictions have been compared with those 
of the \mcatnlo{} and \powheg{} methods. 
A good agreement, within the NLO accuracy, 
has been found between the default options of these methods. 
For $p_T^\text{H} > M_H$ the \krknlo{} result shows better agreement with the 
\powheg{~(Original)} option, a result of the restriction on the transverse momentum 
of parton-shower emissions to below the factorization scale present in the other setups.

Finally, the theoretical predictions of the above NLO-PSMC matching 
methods have been compared with the ATLAS data
from LHC Run~1 for several observables for the Higgs-production process. 
All of the matching methods
underestimate the ATLAS measurements, however the experimental errors are large. 
The \krknlo{} results offer comparable predictions to other matching methods in 
all distributions and, along with \powheg{ (Original)}, 
predict marginally harder spectrum for high $p_T$ and larger jet multiplicities.
It is worth mentioning that all the calculations are performed in the infinite top-quark mass 
approximation, therefore including finite quark mass effects, which are important 
for large transverse momenta, would most likely bring the predictions
closer to the experimental data.

As a next step in our numerical predictions with the \krknlo{} method 
we plan to perform a more detailed study 
of the Drell--Yan processes, 
involving both the neutral ($Z/\gamma^*$) and charged ($W^{\pm}$) modes, 
in presence of experimental cuts and a focus on leptonic observables. 
In order to do this, we need to supplement the NLO-correcting weights with
appropriate spin correlations for vector-boson decay products 
(which is not needed in the case of the scalar Higgs boson).
This can be done rather easily within the \herwig{} PSMC algorithm 
using the method proposed in Ref.~\cite{Seymour:1994we}. 
Future work will seek to apply the \krknlo{} method to other processes investigated 
at the LHC, first of all looking at electroweak-boson pair-production 
($VV$, where $V=\gamma^*,Z,W^+,W^-$) and $V+$jet production.
This would be an important test of feasibility and universality of the method.

Future work will also comprise an appraisal of uncertainties of the
\krknlo{} approach, similar to that of Ref.~\cite{Bellm:2016voq}.  Beyond this, the
natural extension for \krknlo{} is NNLO + NLOPS, where NLOPS is PSMC that
implements the NLO evolution kernels in the fully exclusive form, and thus
provides  the full set of the soft-collinear counter-terms for the hard process.
Ref.~\cite{Jadach:2013dfd} reviews several feasibility studies showing that
construction of such a NLOPS is, in principle, plausible. Moreover, 
the simplifications due to the \krknlo{}  method in achieving NLO+PS will, in our opinion,
be instrumental towards these more ambitions research directions. 

The \krknlo{} project will be available on {\tt hepforge} at 
\url{https://krknlo.hepforge.org}. This will become the home site of the \krknlo{}
development, containing relevant codes and the MC-scheme PDFs as well as set-up 
instructions to facilitate its use within \herwig{}.

\appendix
\section{PDFs in $\msbar$ and MC schemes}
\label{app:pdfs}

\begin{figure}[!ht]
  \begin{center}
    \includegraphics[width=0.95\textwidth]{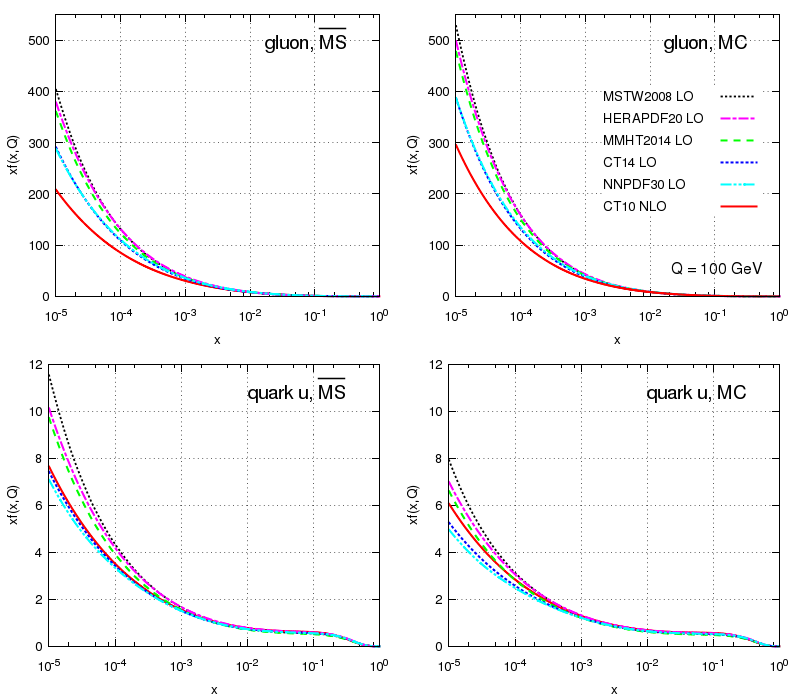}
    \includegraphics[width=0.95\textwidth]{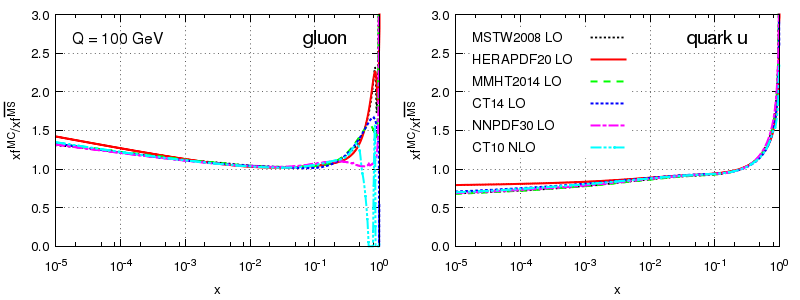}
  \end{center}
  \caption{
    Comparison of different sets of parton distributions functions in the
    $\msbar$ and MC factorization schemes.
  }
  \label{fig:pdfs-comp}
\end{figure}

In this appendix we present comparisons of parton the distribution functions used
in our study. Fig.~\ref{fig:pdfs-comp} shows the $\msbar$ and MC PDFs for the gluon
and for the $u$ quark (remaining quark flavours look very similar). The MC PDFs
were obtained from the $\msbar$ sets using the convolutions discussed in
Ref.~\cite{Jadach:2016acv}.

We observe that the gluon PDF in the MC scheme is larger at small and
moderate~$x$ than the $\msbar$ PDF. On the contrary, the quark PDFs are smaller in
the MC scheme. We also see that various $\msbar$ sets exhibit differences that 
carry on to the MC scheme. These differences lead to a range of predictions
shown in Fig.~\ref{fig:NLO-H-pdf}.


\section*{Acknowledgements}
We are grateful to Simon Pl\"atzer for the useful discussions and his help 
with the dipole parton shower implemented in \herwig{}.
We are indebted to the Cloud Computing for Science and Economy 
project (CC1) at IFJ PAN (POIG 02.03.03-00-033/09-04) in Krak\'ow whose resources 
were used to carry out all the numerical computations for this project. 
We also thank Mariusz Witek and Mi{\l}osz Zdyba{\l} for their help with CC1. 
This work was funded in part by 
the MCnetITN FP7 Marie Curie Initial Training Network PITN-GA-2012-315877.

\providecommand{\href}[2]{#2}\end{document}